\documentclass[preprint2]{aastex7}

\usepackage{orcidlink}
\usepackage{pifont}

\defcitealias{PillsworthRoscoe2025a}{Paper~1}
\defcitealias{PillsworthRoscoe2025b}{Paper~1b}

\begin{document}

\title{Filamentary Hierarchies and Superbubbles II: Impact of superbubbles and galactic dynamics on filament formation and fragmentation}

\correspondingauthor{Rachel Pillsworth}
\email{pillswor@mcmaster.ca}

\author[0000-0002-3033-3426]{Rachel Pillsworth}
\email{pillswor@mcmaster.ca}
\affiliation{Department of Physics and Astronomy, McMaster University, Hamilton, ON L8S 4M1}

\author[0000-0002-7605-2961]{Ralph E. Pudritz}
\email{pudritz@mcmaster.ca}
\affiliation{Department of Physics and Astronomy, McMaster University, Hamilton, ON L8S 4M1}
\affiliation{Origins Institute, McMaster University, Hamilton, ON L8S 4M1\\}

\author[0000-0001-9605-780X]{Eric W. Koch}
\email{}
\affiliation{National Radio Astronomy Observatory, 800 Bradbury SE, Suite 235, Albuquerque, NM 87106, USA}

\author[0000-0003-4852-6485]{Theo J. O'Neill}
\email{theo.oneill@cfa.harvard.edu}
\affiliation{Center for Astrophysics $|$ Harvard \& Smithsonian, 60 Garden St., Cambridge, MA 02138, USA}

\begin{abstract}
    Large scale phenomena in spiral galaxies such as shear, supernovae, and magnetic fields all contribute to the formation and subsequent evolution of filamentary structure and star formation within them. In this paper, we analyze the properties and dynamics of filaments in a simulated Milky Way-like galaxy from \citet{ZhaoPudritz2024}. Using filament and superbubble structure analysis codes, we investigate the roles of galactic shear, supernovae and superbubbles, and magnetic fields on the stability and fragmentation of filaments. We find that local shear has little effect on filament stability and the largest structures at outer radii of the disk may be more likely to be dissipated by shear than supernovae. Filaments are largely parallel to the magnetic field, which plays a significant role in filament stability. By measuring the ratio of surface pressure on a filament to that on its central spine, $\chi_f=P_{surf}/P_{central}$, we find that filaments with  $\chi_f \le 1$ are dominated by their own self gravity and have a strong tendency to be gravitationally supercritical, whereas those with $\chi_f > 1$ are either transitory or in the act of being formed. Finally, we investigate the role of ISM pressure on filament dynamics and stability as a function of galactic radius, finding considerable changes in filament stability and the accompanying star formation rates in the inner versus outer regions of the disk.
\end{abstract}

\section{Introduction}\label{sec:intro}
Physical phenomena such as galactic shear and spiral arms \citep[][]{ElmegreenElmegreen2018, MeidtLeroy2018, ElmegreenElmegreen2019, WilliamsSun2022, QuerejetaLeroy2024, RobinsonWadsley2025}, supernovae and superbubble expansion \citep[][]{McKeeOstriker1977, DobbsBurkert2011, KimOstriker2015, PadoanPan2016, WatkinsKreckel2023, JimenezSilich2024}, and magnetic fields \citep[][]{PlanckCollaborationAdam2016, AbeInoue2021, SetaMcClure-Griffiths2025a, Hu2025} all contribute to the formation of star-forming filaments in the disk in uniquely important ways. The goal of this paper is to use state of the art MHD galactic simulations to better understand the exact role these dynamical processes play in the formation, dynamics, stability, and fragmentation of filaments formed in the global galactic environment undergoing supernova feedback. 

Our previous simulation papers characterized the statistical properties of filaments from tens of pc to kpc scales. In particular, we computed the probability distribution functions for the lengths, masses, and line masses of galactic scale filaments. We found that the most massive filaments in a galactic disk are the longest ones consisting of swaths of a spiral arm \citep[][henceforth referred to as \citetalias{PillsworthRoscoe2025a, PillsworthRoscoe2025b}]{PillsworthRoscoe2025a, PillsworthRoscoe2025b}. 
We also found that spiral arms host the largest scale structures and may therefore be important in the formation of GMCs. Finally, superbubbles were coincident with many of our filaments but we did not then have the analaysis tools to map them and investigate their impact.

The nature of the arms, whether a strong grand-design or a more flocculent structure, could have consequences for the star formation rates of the galaxy. The classification of the Milky Way is a topic of some debate in the current literature, with reasonable arguments of it being a grand-design spiral \citep{ReidMenten2019} or a flocculent spiral \citep{BalserBurton2025}. No matter the strength of its arms,\textbf{ }observations show that the spiral arm segments host most of the Milky Way's star formation \citep{ColomboDuarte-Cabral2022, ColomboKonig2021, SchullerUrquhart2021, Roman-DuvalJackson2010}. Extragalactic observations \citep{ElmegreenElmegreen2019, FinnJohnson2024, QuerejetaLeroy2024} also suggest that spiral arms are the largest scale of star formation, hosting higher fractions of massive star-forming regions than interarm regions of the disks. 

The spiral arm of a galaxy drives a shock wave in the ISM that sweeps up and concentrates the gas into filamentary structure. At the same time, this curved shock creates supersonic turbulence \citep{PudritzKevlahan2013} that pervades the forming filament. This naturally implies that the densest gas in a galaxy is collected primarily in the spiral arms \citep{VogelKulkarni1988, HelferThornley2003, ElmegreenElmegreen2018, ColomboDuarte-Cabral2022, FinnJohnson2024}. 

Supernovae power expanding bubbles of hot gas that drive sweeping, compressive shock waves in the ISM, propagating and colliding with one another \citep[as first discussed in the 3-phase ISM model of][]{McKeeOstriker1977}. These compressed cooling shell walls are natural environments for filament formation.

The shearing effect of galactic rotation stretches and deforms the superbubbles \citep{WatkinsKreckel2023}. A supernova bubble can be stretched into a prolate spheroid, with its major axis aligned with dense structures (i.e. filaments, molecular clouds) in the gas \citep{RomanoBehrendt2025, WatkinsBarnes2023}, suggesting a connection to the shear in the disk. \citet{RomanoBehrendt2025} also find superbubble deformation to occur sooner when nearby dense filaments or other anisotropies, suggesting that superbubble structure can be influenced by a filamentary network. 

All of these considerations motivate this paper wherein we develop techniques to investigate the roles galactic shear, superbubbles, and ISM pressures in the formation and stability of filaments. We extend the analysis from \citetalias{PillsworthRoscoe2025a} in several ways. In order to differentiate superbubbles in our galactic disk, we implement a new bubble segmentation code \texttt{perch} (O'Neill et al., in prep.) and outline our data and structure analysis methods in \S \ref{sec:methods}. The identification of the bubbles is greatly affected by the hot gas included in our projection. This motivates \S \ref{sec:prep} where we compute the impact of different choices of projection depths and observationally motivated column density thresholds on the analysis of the simulation data. The results  of our final projection choice are given in \S \ref{sec:temp} and \ref{sec:linemass}. We then present our physical results in \S \ref{discussion}, where we investigate the importance of dynamical processes such as galactic shear, superbubble proximity, and pressure (external and internal to filaments) on filament structure and stability.

\section{Simulations \& Methods}\label{sec:methods}
The numerical data that we use in this paper come from the simulations of a Milky Way type galaxy in \citet{ZhaoPudritz2024}, which are run in \textsc{ramses} with the AGORA project initial conditions \citep{KimAgertz2016}. These include a dark matter halo with M\textsubscript{DM halo} = 1.074 x $10^{12}$ M\textsubscript{\(\odot\)}, R\textsubscript{DM halo} = 205.5 kpc, and a circular velocity of v\textsubscript{c,DM halo} = 150 km/s, an exponential disk with M\textsubscript{disk} = 4.297 x $10^{10}$ M\textsubscript{\(\odot\)}, and a stellar bulge with M\textsubscript{bulge} = 4.297 x $10^{10}$ M\textsubscript{\(\odot\)} (\cite{KimAgertz2016}). The full details of the simulation setup are outlined in \citet{ZhaoPudritz2024}. 

In brief, this simulation contains magnetic fields, supernova feedback and star particles to model the evolution of structure formation in the Milky Way. We note that these simulations do not include expanding HII regions which are likely the most important mechanism of GMC destruction\citep{HollenbachParravano2026}. Additionally, these simulations offer a high resolution (5.2 pc) to analyze the large scale filamentary structure, leaving the analysis of the very high resolution, 3 kpc zoom-in regions of \citet{ZhaoPudritz2024} to a separate, future work. However, the presence of some of our large scale filaments in those zoom regions does show that the process continues to smaller scales and connects the filamentary hierarchy.

We employ the full galaxy simulation data analyzed in the same manner as in \citetalias{PillsworthRoscoe2025a, PillsworthRoscoe2025b}. That is, we take the galaxy at an age of 283.7 Myr and cut the data to a radius of 13 kpc from the center, smoothed to a constant resolution of $\sim5.2~\rm{pc}$ to mitigate effects from the adaptive mesh on our structure analysis. This particular snapshot is from the first star formation epoch of our galactic disk after the simulation has settled into a steady state, with a star formation efficiency of approximately 10\%.

\subsection{Filamentary structure analysis}
We analyze filamentary structures in a 2D column density map of our simulated galaxy with the FilFinder code \citep{KochRosolowsky2015}. In this work, we make slight changes to the preparation of our data before passing to FilFinder which we will outline in the following section \S\ref{sec:prep}. Otherwise, the filament identification parameters remain the same as those outlined in \citetalias{PillsworthRoscoe2025a}. 

We used a global density threshold for masking at a value of $0.003$ g cm$^{-2}$ ($\sim 1.8\times10^{21}$ cm$^{-2}$), corresponding to the 95th density percentile. The masking step of \texttt{FilFinder} performs an initial cut for structures with a minimum 3:1 aspect ratio, and therefore traces the filamentary structure well across a full galactic disk. In the skeletonization and structure pruning step, we set filament spines to a minimum of 5 pixels, representing a physical length of approximately 25 pc and branches to a minimum of 3 pixels, representing a physical length of approximately 15 pc. Widths are measured by a fit to a Gaussian, and given as the FWHM of the curve. In order to ensure we analyze only filaments which can be reasonably described as resolved, we remove any identified filaments that are less than 25 pc from our data. 

\subsection{Superbubble structure analysis}\label{sec:perch}
In this work we use the in-development python library \texttt{perch} (O'Neill et al., in preparation) for superbubble identification\footnote{\url{https://github.com/theo-oneill/perch}} in our data. This software allows us to analyze superbubble properties in our disk as well as remove superbubble shells from the data so that they are not misidentified by \texttt{FilFinder} as filaments. 

Basically, \texttt{perch} applies persistent homology, a technique developed in the field of topological data analysis, to identify and segment voids of different dimensions in an image or data cube.  The 0$^{\rm{th}}$ homology group corresponds to connected components, while the 1$^{\rm{st}}$ homology group corresponds to loops enclosing lower-valued voids (such as, e.g. the dense walls of a superbubble surrounding the low-density, cleared-out cavity from the explosion). In the context of our 2D column density map, the 0$^{\rm{th}}$ homology group includes higher-density connected structures such as filaments or clouds (similar to structures identified with dendrogram structure finding or \texttt{FilFinder}), while the 1$^{\rm{st}}$ homology group includes lower-density structures such as bubbles, inter-arm regions, and other generic voids in our galactic disk.  If instead applied to our 2D temperature map, the 0$^{\rm{th}}$ homology group includes higher-temperature connected structures such as hot bubbles that have experienced recent feedback.

\texttt{perch} segments and hierarchically maps structures in each homology group that are deemed ``significant'' by the user, such that each individual significant bubble or filament can be given a unique identifier in a structure list and can be filtered for in analysis steps.  We define a significance threshold as a minimum persistence value of 0.5, normalized by the structure's highest valued cell -- its birth -- in\textbf{ }the homology group of interest. We refer the reader to O'Neill et al., in preparation, for further details on the methods of \texttt{perch}.

We use \texttt{perch} to identify superbubbles in our galactic disk in two ways. First, we use the column density map alone to identify all 1-dimensional voids in our galactic disk. For the purposes of the analysis in \S \ref{sec:pressure}, this is the only step in our ``bubble'' structure identification as we care to identify both low-density voids (either older, cooled superbubbles or other dynamical processes creating holes) and superbubbles (younger, filled with hot gas) with no need to distinguish between the two at this step. We additionally compute a second map using the temperature map of our simulated galaxy to identify hot 0-dimensional voids, representing hot bubbles. 
We create a joint mask of the pixels identified in both the density map's 1st homology group and the temperature map's 0th homology group to limit our bubble candidates to only the hot, actively expanding bubbles\footnote{We note that the expansion of the bubbles is verified by eye after structure segmentation, and is not an explicit parameter in the \texttt{perch} analysis.}. This combined mask is what we use for our data preparation, in order to filter out the superbubble walls from our filament search. We also use it to investigate the temperature structure of our simulated galaxy in \S \ref{sec:temp}. We do not use the 0th order homology map of our density for any filamentary structure analysis. We restrict our filament data to only that produced via \texttt{FilFinder}. 

\begin{figure*}
    \centering
    \includegraphics[width=0.85\linewidth]{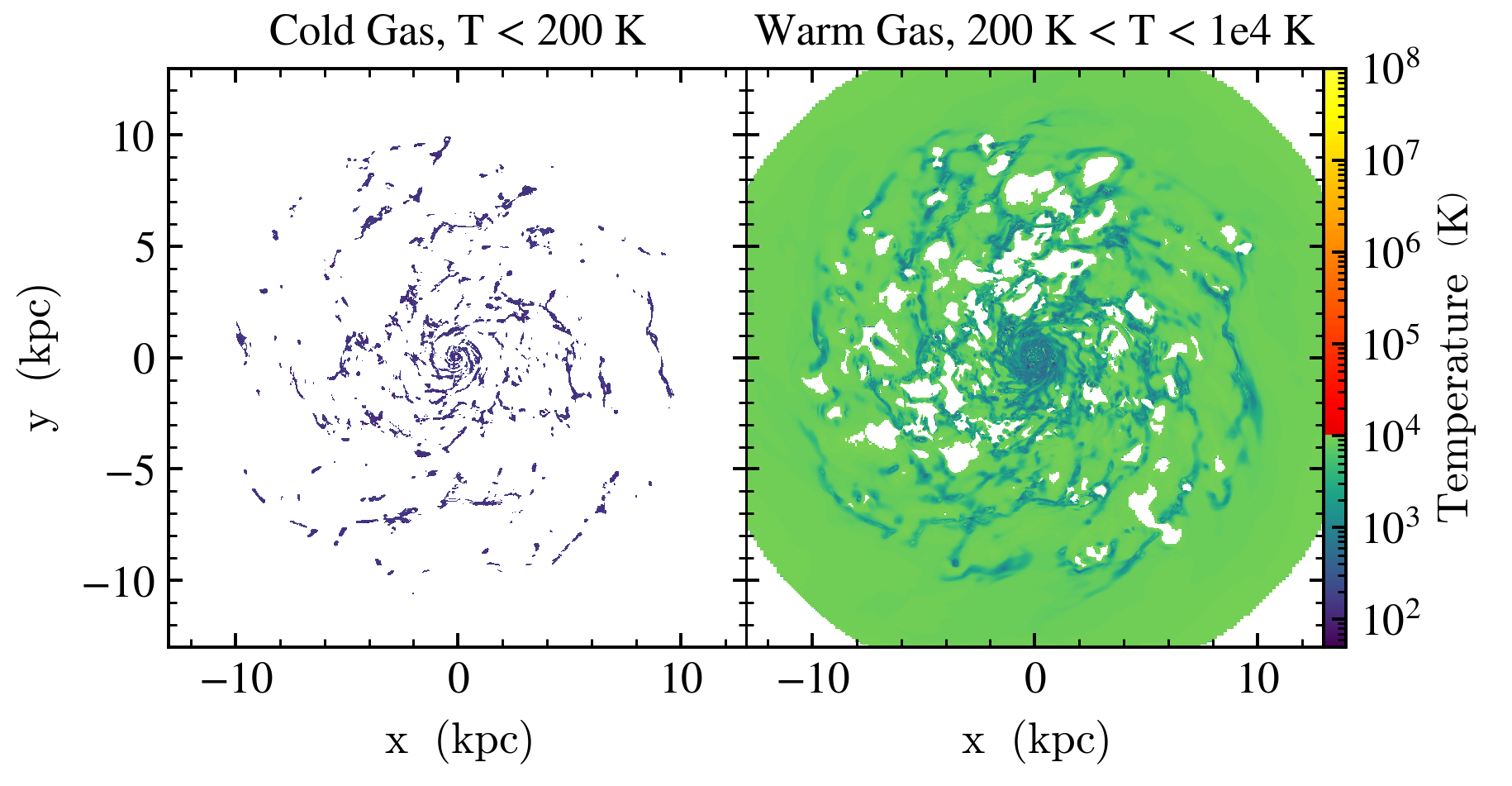}
    \includegraphics[width=0.86\linewidth]{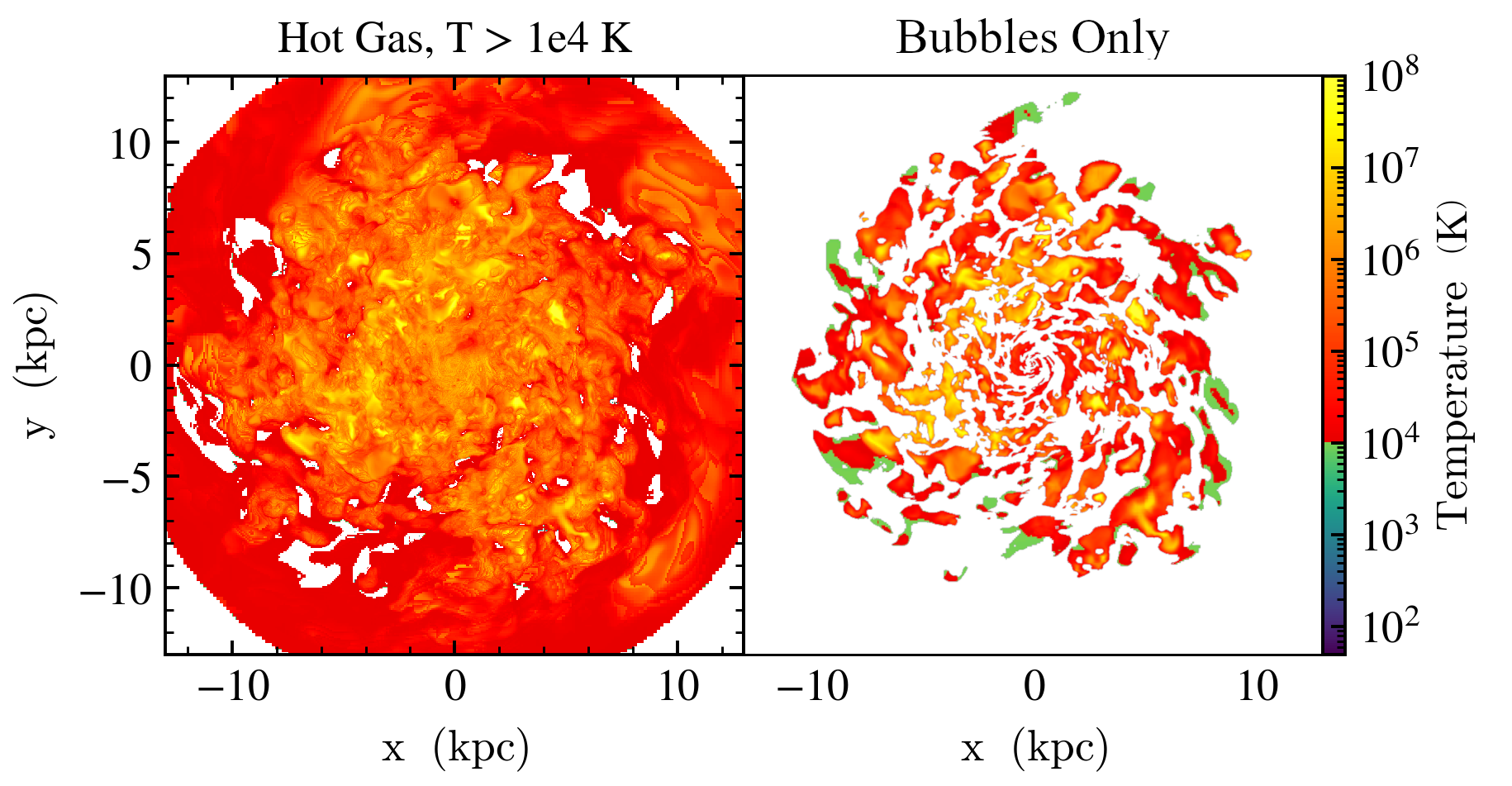}
    \caption{Different temperature projection cuts of the gas in the galaxy from our pillbox projection. \textit{Top row:}From left to right, we show cold gas ($<$200 K) with filaments overlaid and the warm gas between 200K and 10$^4$K with filaments overlaid. \textit{Bottom row:} From left to right, we show all the gas hotter than $10^4$K, and the hot gas in the bubbles from \texttt{perch}.}
    \label{fig:temp}
\end{figure*}

In Figure \ref{fig:temp} we show the projected temperature maps for the cold gas below 200K and the warm gas between 200K and $10^4$K, both of which showcase the gas $\sim1000$ K and below which make up the filamentary structure. In the bottom row of the same figure, we show the hot gas above $10^4$K and the gas contained in our superbubbles identified by \texttt{perch}. Further discussion of this figure is presented in \S \ref{sec:temp}.

\subsection{Critical line masses and definitions}
In \citetalias{PillsworthRoscoe2025a}, we provide an overview of the critical line mass calculations of filaments, and the subsequent definitions of super- and sub-critical filaments. In brief, the critical line mass for filaments with only thermal gas motion was derived by \citet{Ostriker1964}; $m_{crit,thermal}=2\sigma^2/G$ where $\sigma^2= c_s^2$. This thermal critical line mass neglects the turbulent motions, but in the current literature these are most often included in the form of a velocity dispersion such that $\sigma^2 = c_s^2 + \sigma_{NT}^2$ to define a total velocity dispersion \citep{FiegePudritz2000} -- so named because it is the standard convention of observational work to measure only thermal and non-thermal turbulent motions \citep{HacarClark2023, SyedSoler2022, JacksonFinn2010}. In addition to this ``total" velocity dispersion, one can include magnetic field effects by the addition of Alfv\`{e}n speed as in \citet{FiegePudritz2000} and retrieve a magnetic critical line mass.

\begin{equation}\label{eq:clinemass}
    m_{crit} = 2(c_s^2 + \sigma_{NT}^2 + v_A^2)/G
\end{equation}
\noindent All of these contributions are evaluated in the different local environments in our galaxy simulation, and were considered in the filament criticality analysis of \citetalias{PillsworthRoscoe2025a}. 

This critical line mass can be extended to include shear effects. This can be seen in principle by noting that the critical line mass can be derived from a virial analysis of magnetized filaments in a turbulent medium that includes thermal pressure \citep{FiegePudritz2000}. Since a virial analysis includes all relevant energy densities, it is possible to extend it to include the energy density in the shear field. In this work, those shear effects arise as an effective shear dispersion $\sigma_{sh}$. 

We investigate the role of streaming motions by investigating the local shear of the gas. From first principles, shear is the symmetric component of the strain tensor. Therefore, in a simple 2D, Cartesian case, one can define the shear as

\begin{equation}
   \Omega = |\frac{dv_x}{dy} + \frac{dv_y}{dx}|
\end{equation}

\noindent which can be turned into a shear velocity dispersion with $\sigma_{sh} = \Omega \times dr$, where dr is the spatial resolution across which the shear is sampled -- in our case $5.2~\rm{pc}$. For purely rotational cases, the shear represents the differences in rotation speeds at different radii, creating a shearing zone which may serve to add to rotation of a structure located within it. For galaxies, the rotational shear is increased by the spiral arms sweeping through the disk. On smaller kpc or 100 pc scales, supernova explosions can increase non-rotational shear. 

In our calculations of shear we compute the 3-dimensional local shear map throughout our disk\footnote{This is computed through a derived field defined in the YT-project analysis code \citep{TurkSmith2011}.} and project it as we do the rest of our data into a 2D, density-weighted map. This projected local shear map is multiplied by our cell size of 5.2 pc to produce local shear dispersion measurements.

Thus, the expression for the critical line mass in this work is made more general by adding the galactic shear in quadrature. A full consideration of the critical line mass of a filament in this work is

\begin{equation}
    m_{crit,sh} = 2(c_s^2 + \sigma_{NT}^2 + v_A^2 + \sigma_{sh}^2)/G
\end{equation}
\noindent where each individual term represents a gas motion due to a large-scale galactic process. We note here that the ``total" velocity dispersion still remains to be only the thermal and turbulent contributions, whereas magnetic fields and galactic shear are specifically not turbulent processes. 

The ratio of the actual measured 1D density, or line mass, of a filament to this critical line mass determines whether it is supercritical ($m/m_{crit}>1$) or subcritical ($m/m_{crit}<1$). Throughout this paper we refer to the line mass ratio as $f_x=m/m_{crit,x}$, where x denotes which version of the critical line mass we use. For instance, including only thermal effects is denoted as the thermal line mass ratio, $f_{therm}$. Importantly, definitions of critical line masses and criticality are additive. The included processes are added on to the previous definition. Note that we investigate more processes than is usual in the literature. Whereas many papers discuss the combination of thermal and turbulent effects as the ``total" \citep[such as in][]{HacarClark2023}, we denote these thermal and turbulent contributions as $t+NT$ - which is just one important component of the total contributions possible for the critical line mass. This change in subscript clarifies what contributions to critical line mass are considered. We summarize these definitions in Table \ref{tab:defs} for clarity and ease of reference, where one can see that, for instance, the shear critical line mass is the magnetic critical line mass with the addition in quadrature of the shear dispersion.

\begin{table*}
    \centering
    \begin{tabular}{cccc}
        \textbf{Term used in text} & \textbf{Critical line mass} & \textbf{Line mass ratio term} \\ \hline
        Thermal critical & $m_{crit, therm} = 2c_s^2/G$ & $f_{therm} = \frac{m_{line}}{m_{crit,therm}}$ \\
        Turbulent critical & $m_{crit, t+NT} = 2(c_s^2 + \sigma_{NT}^2)/G$ & $f_{t+NT} = \frac{m_{line}}{m_{crit, t+NT}}$\\
        Magnetic critical & $m_{crit, b} = 2(c_s^2 + \sigma_{NT}^2 + v_A^2)/G$ & $f_{b} = \frac{m_{line}}{m_{crit, b}}$\\
        Shear critical & $m_{crit,sh} = 2(c_s^2 + \sigma_{NT}^2 + v_A^2 + \sigma_{sh}^2)/G$ & $f_{sh} = \frac{m_{line}}{m_{crit, sh}}$ \\
    \end{tabular}
    \caption{Critical line mass names, equations, and line mass ratio used throughout this work.}
    \label{tab:defs}
\end{table*}

\section{Impacts of Preparation and Projection of Data}\label{sec:prep}
As already indicated,  \texttt{perch} allows us to analyze superbubbles as individual, hierarchical structures in our data. However, hot gas within the superbubbles in these global galactic simulations is usually ejected into the galactic halo. This means that there is a substantial amount of hot gas above and below the galactic disk in our simulation which affects the accuracy of perch's superbubble segmentation in the galactic plane when using face-on views of the simulated galaxy. It is therefore important to remove the hot halo gas component. 

Our original method in \citetalias{PillsworthRoscoe2025a} used a full-box projection to produce a 2D column-density map. This procedure includes all of the halo gas and rasies both the column density and total temperature in our projection, preventing accurate bubble identification with \texttt{perch}. Accordingly, we investigate the effects of projecting our data along a shorter line-of-sight and compare this to constraints posed by observational methods. Our technique has two benefits: it enables our software to more clearly identify superbubbles while also providing an approach to preparing simulation data which can be compared to observational constraints without losing our kpc scale filament population.

With respect to the latter point, observations recover only a limited amount of information about gas properties in the ISM compared to simulations. For example, commonly-observed spectral lines such as $^{12}$CO, $^{13}$CO, HCN all trace the molecular gas in the ISM, sensitive to specific ranges in gas density and temperature. Other tracers, such as dust extinction and emission, can trace the neutral ISM more broadly but are subject to their own limitations, including empirical calibrations of the gas-to-dust ratio, dust opacity, and the interstellar extinction curve.

Simulations, in contrast, most often treat gas as a single fluid. Although observations of multiple gas phases often require multiple observing cycles for different tracers --- many of which may not be available at similar resolutions --- simulations are always able to work with subsets of all of the information at the same resolutions. In fact, the specific strength of simulations is that all the gas data is available at once from one simulation. As such, in this section we highlight the benefits of methods which preserve this strength by limiting domains and categorizing structures.

First, we carry out projections across our full simulation domain (the ``original" data preparation method of \citetalias{PillsworthRoscoe2025a}) and projections limited to a depth of $\pm427$ pc from the midplane (the ``pillbox" data, so named as it is limited to a squat cylinder centered on the galactic midplane). We choose $\pm427$ pc as it both represents the height at which our radially-averaged galactic pressure has dropped from the average midplane pressure by one e-fold, representing a hydrostatic scale height; and it encompasses the entirety of the Milky Way's observed thin disk \citep[$\approx300$ pc at the Solar neighborhood, depending on the observable][]{Bland-HawthornGerhard2016} and average HI scale heights from cosmological zoom simulations \citep{GensiorFeldmann2023}. We do not account for any expected flaring of the disk in our pressure scale height measurement. Due to limitations in the allowed geometry for projections of 3D data, we take the radially-average galactic pressure, out to a radius of 13 kpc, and measure one pressure scale-height across this data.

For an observational comparison with recent dust studies, we investigate the filament population with column density cuts motivated by comparisons to cuts from recent 3D dust mapping analyses by \citet{ONeillZucker2024}. In particular, we choose column density thresholds of $0.012~\rm{g~cm}^{-2}\sim 7 \times 10^{21}~\rm{cm}^{-2}$ (henceforth the `low' case) and $0.029~\rm{g~cm}^{-2}\sim 10^{22}~\rm{cm}^{-2}$ (henceforth the `high' case) which correspond approximately to $A_G=3$ and $A_G=7$, respectively, assuming a column density conversion of $A_G/N_H \simeq 4.0 \times 10^{-22}~\textrm{cm}^{2}~\textrm{mag}$ \citep{ONeillZucker2024}. These choices correspond to typical column densities found in the CNM and those of molecular clouds, respectively \citep{PillsworthPudritz2024b}. We also note that these cuts are similar to observational findings for the two regimes of magnetic field orientation and, therefore, different degrees of magnetic field support, in the filaments \citep{PlanckCollaborationAdam2016}. 

Finally, we also perform a pseudo-column density cut by masking superbubble structures with the \texttt{perch} code as detailed in \S \ref{sec:perch}. Table \ref{table1} summarizes these different methods of data preparation. The column density cuts are performed on the pillbox projected data without the superbubble masking, representing a map most similar to observations.

\begin{table*}
    \centering
    \begin{tabular}{ccccc}
       \textbf{Dataset}  & \textbf{Pillbox ($\pm$427 pc)?} & \textbf{Column density cut?} & \textbf{Superbubble mask?} \\ \hline
        original & \ding{55} & \ding{55} & \ding{55}\\
        pillbox & \ding{51} & \ding{55} & \ding{55}\\
        perch & \ding{51} & \ding{55} & \ding{51}\\
        low & \ding{51} & $\sim 7 \times 10^{21}~\rm{cm}^{-2}$ & \ding{55}\\
        high & \ding{51} & $\sim 10^{22}~\rm{cm}^{-2}$ & \ding{55}\\
    \end{tabular}
    \caption{Outline of all the considerations in data preparation before running \texttt{FilFinder} analysis. From left to right we consider a projection limited to $\pm 427$pc, corresponding to the approximate pressure scale height of the thick disk in our model; we consider column density cuts motivated by dust extinction values explored in \citet{PillsworthPudritz2024b,PlanckCollaborationAdam2016}; and we consider the superbubble structure mask.}
    \label{table1}
\end{table*}

\subsection{Column density thresholds}
\begin{figure*}
    \centering
    \includegraphics[width=1.0\linewidth]{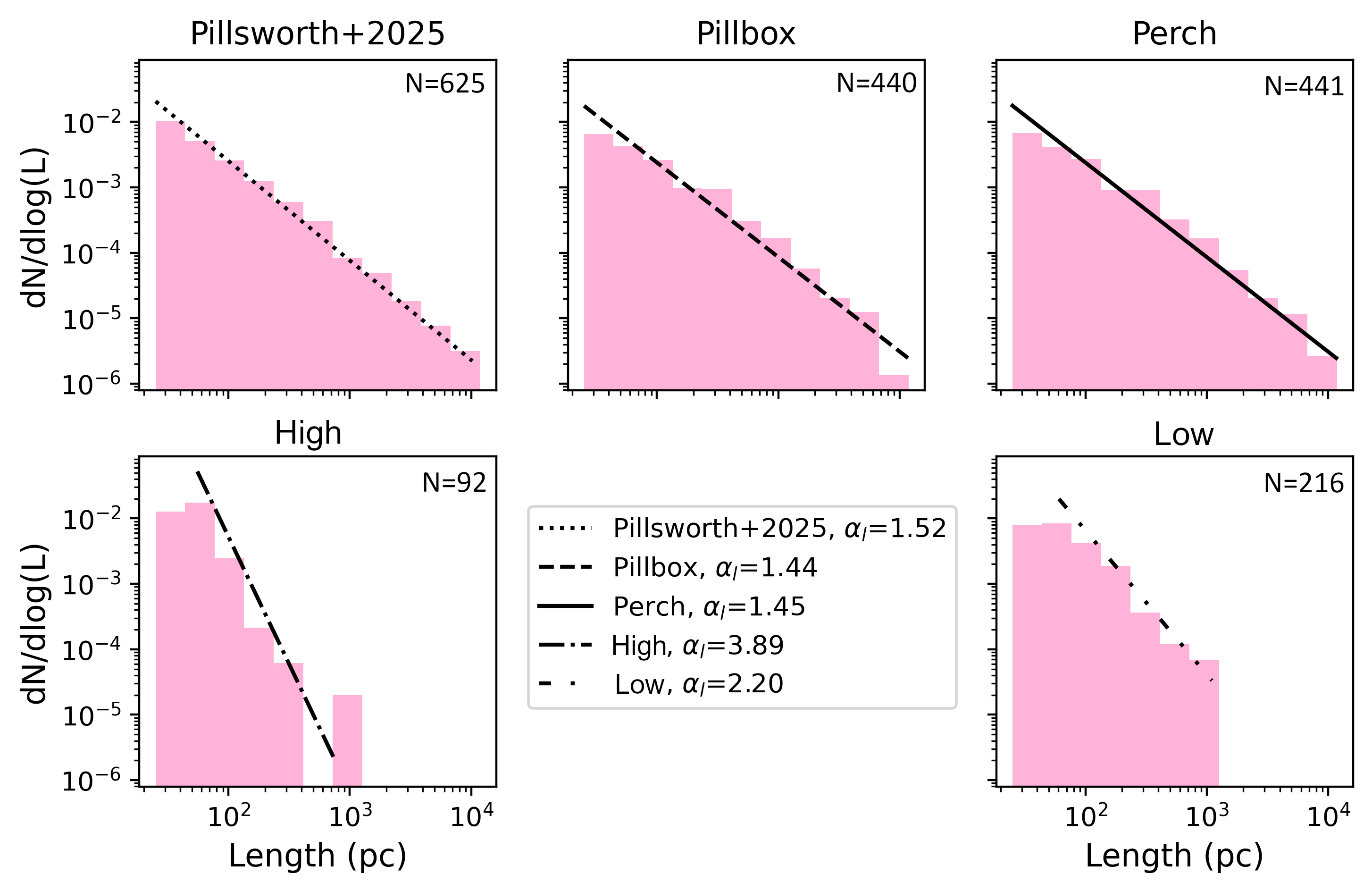}
    \caption{Length PDFs of the filaments identified from different preparations of our galaxy simulation data. The top left panel shows our original filament population from \citetalias{PillsworthRoscoe2025a}, top center shows the filaments identified in a projection limited to distances of +/-427 pc of the midplane (pillbox projection) of the galaxy and the top right distribution shows the filaments identified in the pillbox projection with hot superbubbles masked out of the data. Bottom left and bottom right panels show the distributions for our column density cuts of $\sim10^{22}~\rm{cm}^{-2}$ (\textit{titled High}) and $\sim7\times10^{21}~\rm{cm}^{-2}$ (\textit{titled Low}). The power-law fits for each distribution are plotted in black.}
    \label{fig:lengths_proj}
\end{figure*}

\begin{figure*}
    \centering
    \includegraphics[width=1.0\linewidth]{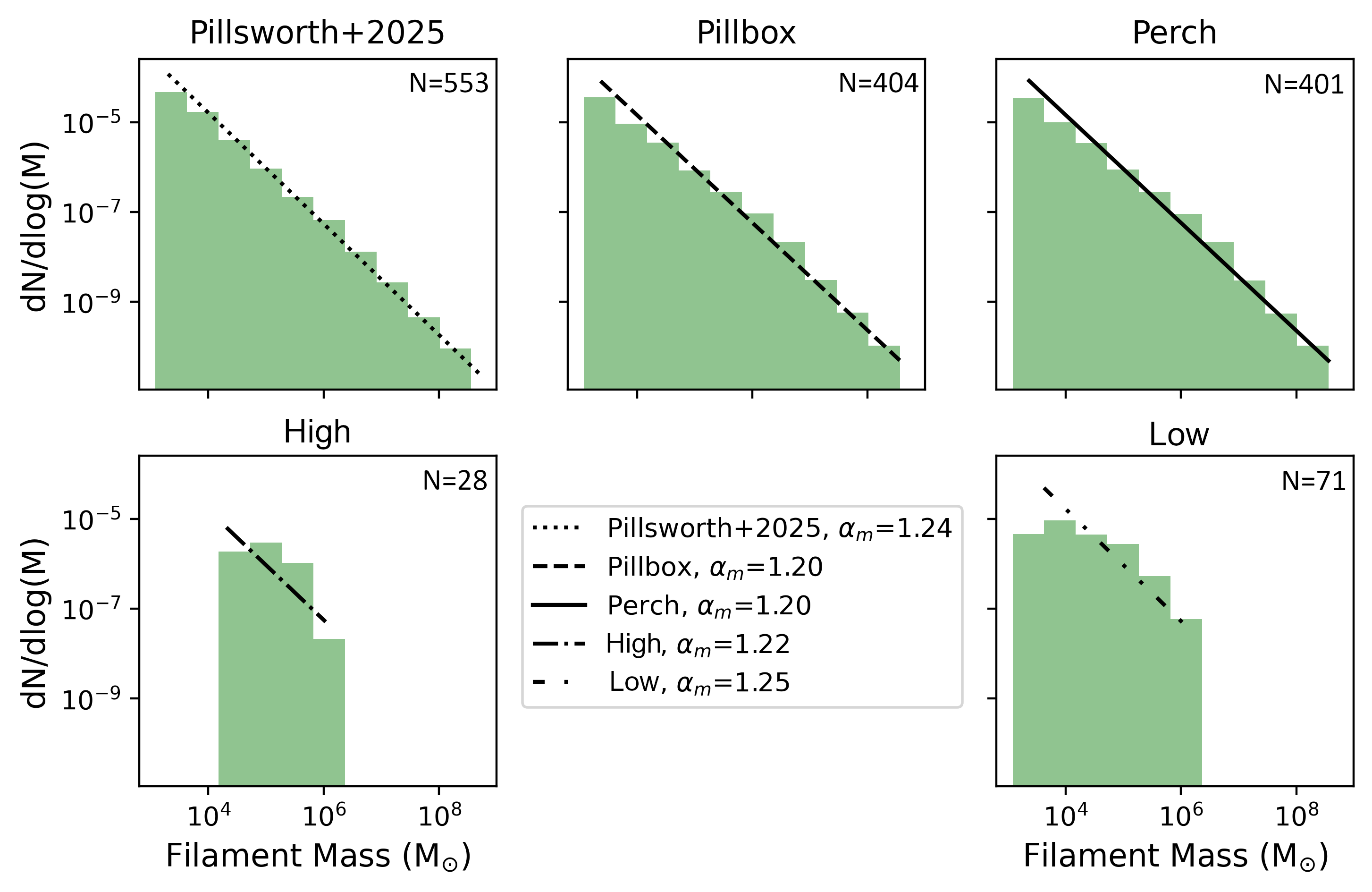}    
    \caption{Mass PDFs of the filaments identified from different preparations of simulation data. Top left are the ``original" filaments from \citetalias{PillsworthRoscoe2025b}. The top middle panel is the distribution from the pillbox projection, while the top right shows the distribution of the pillbox projection with hot superbubbles masked out. Bottom left and bottom right panels show the mass distributions for our column density cuts of $\sim10^{22}~\rm{cm}^{-2}$ (\textit{titled High}) and $\sim7\times10^{21}~\rm{cm}^{-2}$ (\textit{titled Low}).}
    \label{fig:masses_proj}
\end{figure*}

In Figures \ref{fig:lengths_proj} \& \ref{fig:masses_proj}, we show the length and mass distributions of the different preparation methods  outlined in Table \ref{table1}. Compared to the ``original" filament distributions using the methods in \citetalias{PillsworthRoscoe2025a} -- the panels titled Pillsworth+2025 -- the pillbox projection shows little change in the length distribution and virtually none in the mass distribution, except in the total number of filaments identified. The pillbox projection has a reduced number of filaments of lengths 30-50 pc and of 7-10 kpc, especially evident at the truncation of the longest scale filaments. However, there is only a minimal difference between the power-law slopes of these distributions. 

Masking out gas data from superbubbles minimally changes the distributions from the pillbox projection as we recover the same power law index from both distributions. Instead, we find the greatest change in the distribution when considering observationally-motivated column density cuts, the high and low cases in the bottom row of both the length and mass distribution panels. 

Applying different column density cuts to the simulation data results in different power-law fits to the length distributions. These steeper profiles come from the sharp fall-off at lengths of 1000 pc. With column density cutoffs similar to typical dust extinction values for filament observations, no filaments greater than 1 kpc remain. Thus the entire kpc scale of the filamentary hierarchy present in the simulation would not be detectable at this observational sensitivity. This threshold in column density pertains to limits in filament length in current observations, such as those in \citet{HacarClark2023}. 

The mass distributions behave similarly; mass in the simulations is undetectable or ``missing" to varying degrees depending on the observational column density cut that is applied to the numerical data. As an example, our `low case' fails to recover the highest mass filaments, just as it fails to recover long filaments. As such, observational column density cuts corresponding to $\sim7\times10^{21}~\rm{cm}^{-2}$ cannot trace coherent structures in HI-like gas that are produced in our galactic simulations and therefore miss all massive kpc filaments we find that trace the spiral arms. The mass distribution for the high case returns no high mass ($>2\times10^6~\rm{M}_{\odot}$), or even low mass ($<10^4~\rm{M}_{\odot}$), filaments in the distribution, indicating two issues. The first issue pertains to the low mass filaments. These structures are likely cut as a consequence of their lower density gas falling below the threshold cut. Meanwhile, the high mass filaments are made up of the longest filaments. The column density cut will remove the lower column density gas and thereby reduce the apparent connectivity of these filaments. 

Filaments measured from the `low' and `high' column density cuts recover mass distributions that cover barely 2 out of the total 5 orders of magnitude in mass that the filaments cover without these cuts.  Thus, we have shown in these examples how observational limitations in detecting low column density gas affect the extent to which structures -- as they appear in simulations -- can be discerned.

\subsection{Projection effects}
Compared to the column density cuts, the effects of projection depth are more subtle. Here we consider the effect of contributions to the projected column density from material beyond the pillbox cuts and highlight the effect of the projection depth on the line mass ratio of the filaments.

Although we use a density-weighted projection in all cases (Table \ref{table1}), the hot and warm gas beyond $\pm432$~pc of the galactic disk affects the projected filament properties we recover. Within our pillbox projection, the average column density is $\sim10^{-2}~\rm{g~cm}^{-2}$ ($\sim10^{21}~\rm{cm}^{-2}$) outside of superbubbles and gas is sitting at temperatures of $10^4$K or lower (i.e., the neutral phases). In the region outside the pillbox, which was included in \citetalias{PillsworthRoscoe2025a}, the temperature has a significant contribution from gas $\sim5000-8000$K (the warm, neutral phase of the ISM) across the area of the galactic disk so that the average column density is still $\sim10^{-3}~\rm{g~cm}^{-2}$. This warm neutral and hot ionized gas from outside the pillbox, which is not part of any filamentary structure, may serve to blur the structures in the original projection. In addition, this gas contributes to higher apparent levels of thermal support across the shorter, smaller scale filaments. This blurring increases the column densities of filaments and affects both the length and width calculations of filamentary structures in \texttt{FilFinder}, as both rely on drop-offs in density to determine structure edges. These increased lengths and widths of filamentary structure result in apparent steeper length and mass distributions as longer, but not significantly more massive, filaments become more abundant. 

\begin{figure}
    \centering
    \includegraphics[width=1.0\linewidth]{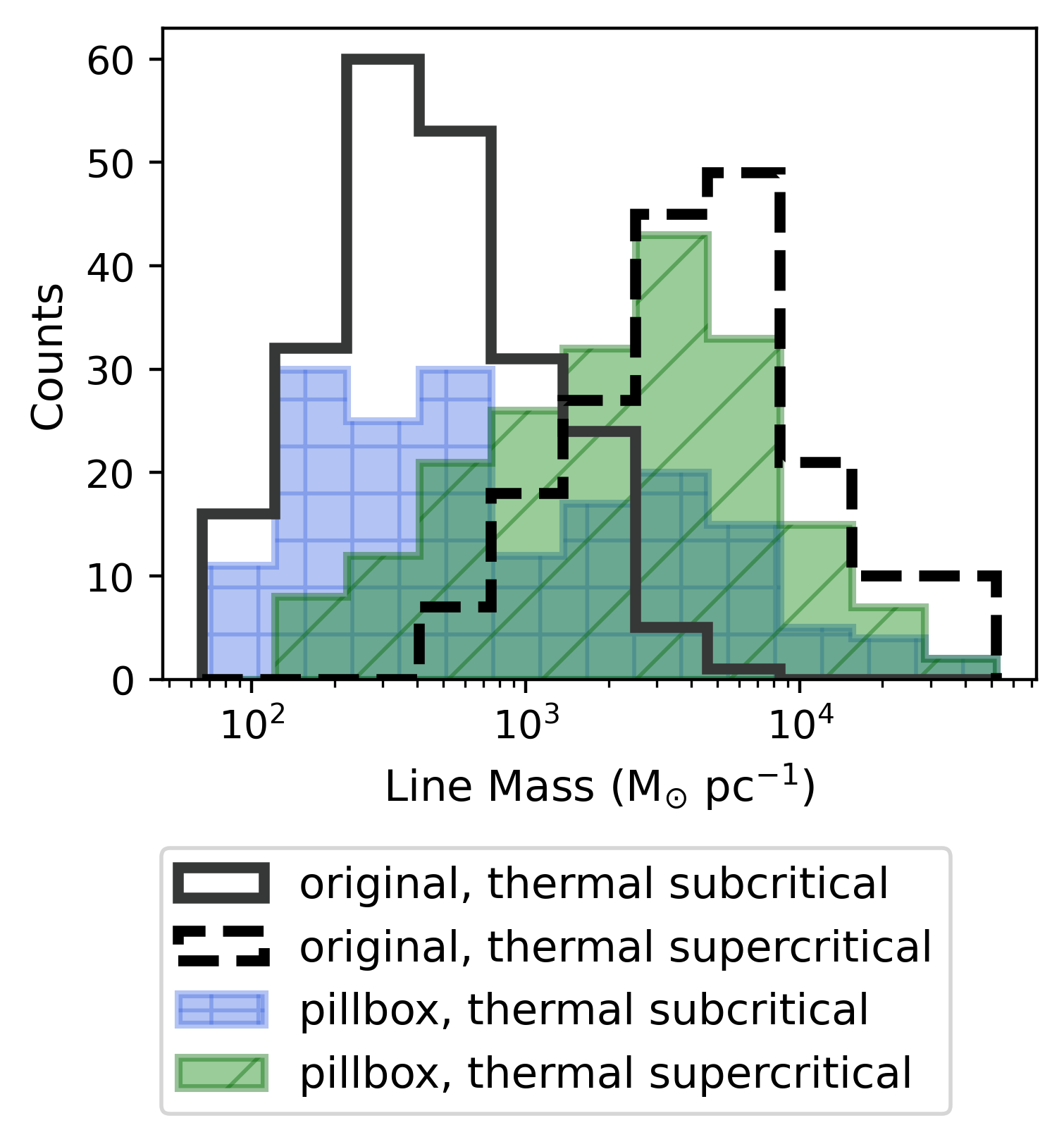}
    \caption{Comparison of our original filament population's line masses \citepalias{PillsworthRoscoe2025a, PillsworthRoscoe2025b} with the filaments from our pillbox projection. In the case of the original population, we show subcritical filaments as the grey line and supercritical as the black, dashed line. For the pillbox case, we show subcritical filaments in blue and supercritical filaments in green.}
    \label{fig:m-l-comparison}
\end{figure}

In Figure \ref{fig:m-l-comparison}, we show the effects of both full-depth and pillbox projections on the inferred line mass ratio of the filaments with data from \citetalias{PillsworthRoscoe2025b} and the \texttt{perch} case from this work. Each of these filament populations are separated into their subcritical and supercritical cases. While their masses and lengths show little to no difference (as seen in Figures \ref{fig:lengths_proj} \& \ref{fig:masses_proj}), we find differences in the line masses measured. 

In the low line mass regime of this Figure, the pillbox method registers more thermally supercritical filaments for lower measured line masses than the original method. Thus, there is an apparent decrease in the thermal support of filaments which are also preferentially our shortest scale filaments. The decreased thermal support that arises from the pillbox method shows that a limited projection depth is most important for shorter scale, and therefore likely cooler, filaments. The importance of the depth lies mainly in the contribution of hot and diffuse gas directly above and below a galactic disk to the computed line mass. 

At the high line mass end in Figure \ref{fig:m-l-comparison}, we note a flipped relation. The original projection (i.e., including gas above and below the galactic disk) contains more supercritical filaments at high line masses and, therefore, longer lengths. However, the pillbox projection finds a mix of subcritical and supercritical filaments at high line masses, with the thermally subcritical population extending to higher line masses than the original method's. Therefore, longer filaments, which tend to exist at outer radii, experience higher thermal support. Moreover, the pillbox shows higher temperatures only at outer radii, but the gas above and below the disk has temperatures of $\sim10^7-10^8$K even at inner radii of the disk. This flipped relation from low to high line masses then means that a limited depth projection can also be key in resolving the temperature differential across galactocentric radii by excluding the opposite temperature gradient found outside the pillbox from galactic fountains. 

We note that across the original, pillbox, and perch data projection methods, we recover the same large-scale population of filaments. This implies the difference in the recovered filament properties is indeed from projection depth. Our longest filaments associated with the spiral arms of the disk have a significant amount of warm neutral and hot ionized gas which would contribute substantially to their column densities in our original method removed by considering the pillbox projection. This shift also decreases the line mass ratio, pushing these filaments to subcritical regimes on average. This result is in line with our previous findings from \citetalias{PillsworthRoscoe2025a} that the largest scale structures will be subcritical on average as the areas of fragmentation are highly localized compared to the extended lengths of the structures themselves.

\subsection{Summary of data preparation and projection effects}
 What is the optimal approach, then, to both identifying the superbubbles in our disk and achieving the best compromise between matching observations and preserving valuable multi-phase gas information from simulations? Our analysis identifies two points of particular note in connecting simulation data to observations. 

The first is that in the projection of simulation data prior to analyzing structures, it is important to ensure that only the gas that is truly part of the structures of interest is being measured. The ideal case of course would be full 3-dimensional analysis for these structures, removing the need to consider projection effects altogether. In this work, however, we focus on understanding the projection effects on simulation data to better compare to observations, particularly dust extinction and emission maps. Given this, it is important to analyze simulation data in a way that removes contaminating hot gas from galactic fountains and retains filament information across a broad range of gas densities. Our analysis shows that a pillbox projection for simulation data is suitable to limit the amount of cuts to make to the data, thereby preserving more information, and to identify the filamentary structure more clearly by reducing the contribution from the warm neutral gas above and below the galactic disk.

The second issue is that, due to the column density thresholds that are an unavoidable consequence of observational techniques, the comparison between simulations and observations is not one-to-one. While it is useful for comparison to reduce simulation data to be as close as possible to observations, we have also found in these results that applying observational techniques to simulations can significantly reduce the structures one can find in simulation data. Stringent density cuts, while mimicking observational data well, neglect a subset of structures in an entire decade of mass and length.

We have found that limiting the projection depth and segmenting identifiable structures into individual maps offers close comparison to observational data by removing contaminating gas and more accurately identifying structures of interest while maintaining the full gas data that simulations have access to. Thus, throughout the rest of this paper we exclusively use data from our `perch' case which utilizes both the pillbox projection to exclude the warm neutral and hot ionized gas above and below the galactic disk and a pseudo-column density cut by further masking out hot bubble structures in the disk with \texttt{perch}. This choice allows us to identify superbubble structures for improved filament identification and further analysis while maintaining the large scale structures in which we are primarily interested. From the mass and length distributions, this case offers the best match with our work in \citetalias{PillsworthRoscoe2025a} while also limiting projection effects from our simulation domain.

\section{Results of Data Analysis}\label{results}

\subsection{Temperature structure}\label{sec:temp}
We analyze the phases of the ISM that filaments and superbubbles exist in by plotting the temperature cuts in the pillbox projected data. Returning to Figure \ref{fig:temp}, we show the cuts of cold gas, warm gas and the hot gas in our galactic disk. The filaments exist in cold and warm gas in the galactic disk, at temperatures of 1000 K or lower. On the other hand, from the ``Bubbles Only" panel in Figure \ref{fig:temp}, we can see that the superbubbles are made up exclusively of gas around $10^4$ K or hotter. These distinct temperature ranges help distinguish between the dense shells of superbubbles and the filamentary structure of the disk.

We find that the superbubbles make up the majority of the surface area of the gas in the galactic disk, encompassing a total surface area of 200 kpc$^2$ (30\% of the total disk surface area). On the other hand, filamentary structure fill a much smaller fraction of the gas, with the mask produced by \texttt{FilFinder} having a surface area of 49 kpc$^2$. This represents only 7\% of the total disk surface area. Filaments make up the majority of the mass budget of our simulated galaxy totalling $3.32\times10^{9}~\rm{M}_{\odot}$, 78\% of the gas mass in the galactic disk.

\subsection{Line Masses}\label{sec:linemass}
\begin{figure*}
    \centering
    \includegraphics[width=0.98\linewidth]{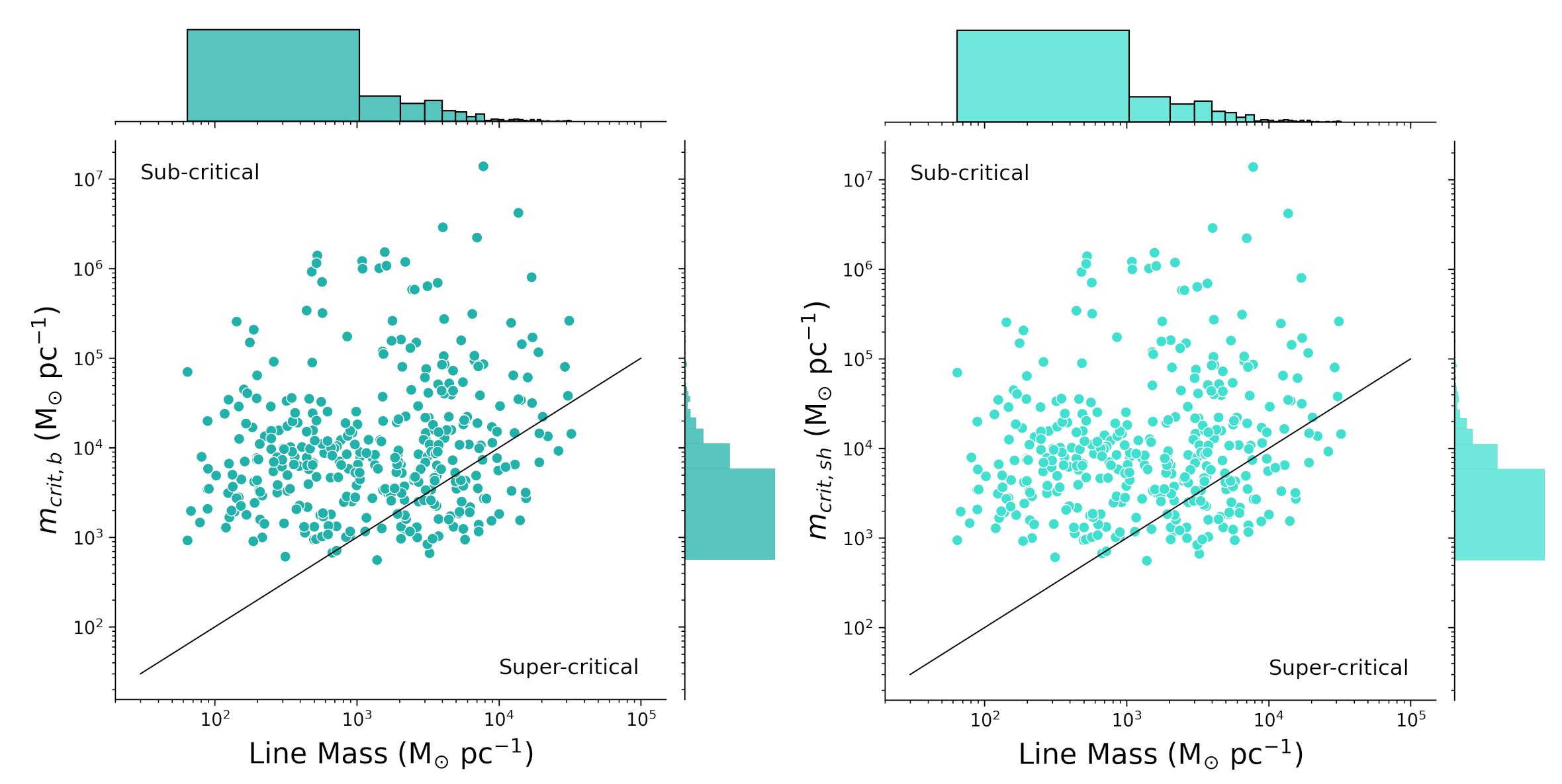}
    \caption{Theoretical critical vs. measured line mass joint plots for the filaments in our population. \textit{Left:} The magnetic critical line mass. \textit{Right:} The shear critical line mass. The black line in each shows the critical ratio of 1, separating the plane into sub-critical and supercritical regions.}
    \label{fig:linemass_scatter}
\end{figure*}

In Figure \ref{fig:linemass_scatter} we plot the values of the magnetic and shear critical line masses for the filaments in our perch case versus their actual measured average line mass.  
We first note the results of including only thermal and turbulent support. Approximately 50\% of our filament population by number (62\% by mass) is thermally supercritical. However, when one factors in the support from internal turbulent motions the proportion of supercritical filaments drops to 39\% by number (50\% by mass). This drop highlights a significant support from internal turbulent motions of the filament. 

More importantly, the addition of support from the magnetic field (left panel) reduces the proportion of supercritical filaments even further, to just 16\% of the filament population by number, but remaining close to half (42\%) by mass. We note that there is no significant drop in magnetic field strength with galactocentric radius. Therefore, the drop we see is not related to outer radii having larger areas for filament formation and much weaker fields. Instead, the significant drop in numbers here highlights the important role that magnetic fields play in supporting filaments against fragmentation, allowing them to reach higher densities before eventually collapsing to form molecular clouds or star clusters. 

The right hand panel shows that even when we consider the shear dispersion in our calculation of the critical line mass of a filament, the population of supercritical structures remains at 16\% by number. Therefore, the local shear around the filament from the galaxy's rotation has little effect in supporting or collapsing the filament. We investigate the difference in the external and internal shearing motions on a filament and discuss in \S \ref{sec:shear}.

Our results agree well with those from \citetalias{PillsworthRoscoe2025b}, showing that when  the support from magnetic fields is included, the percentage of supercritical filaments is in line with star formation efficiencies of 10\% for our galaxy model. Thus even with changes in the projection and preparation of data for structure analysis the degree of magnetic field support still converges. We discuss what the mass percentages of supercritical filaments may imply for these structures in \S \ref{sec:shear}.

Overall, we show that our filament population agrees both with our previous study \citepalias{PillsworthRoscoe2025b} and therefore with observational data from multiple sources \citep{HacarClark2023, SyedSoler2022, VeenaSchilke2021, SchisanoMolinari2020, GoodmanAlves2014, JacksonFinn2010}. The most important consequence is that using only the total average critical line masses would show that many filaments are thermally unstable. This would result in unphysically high star formation rates in the galaxy. It is the inclusion of magnetic field in the critical line mass that aligns criteria for filament fragmentation with star formation efficiencies that are consistent with observed star-formation rates in the galactic disk. Moreover, there appears to be little significant contribution on average from galactic shear to the stability of our filaments. We investigate the possible role both shear and the superbubbles in the disk may play on other properties of the filaments in the following section.

\section{Dynamical Effects}\label{discussion}
We now turn to investigating the roles of various galactic dynamical processes on filament properties and fragmentation. In particular, the shearing motions of the disk, the morphology of the velocity flows, the galactocentric radius at which a filament is positioned, the proximity of superbubbles and the pressures of the filaments compared to galactic values.

\subsection{Shear and filament stability}\label{sec:shear}
The relative roles of shear and compression have been summarized by \citet{AbeInoue2021}. In their terminology, Type S filament formation involves the stretching of structures caused by shearing motions in the disk. Type S filaments often have small line-masses and parallel alignments to the magnetic fields, which makes them much more stable than other filament types. On the other hand, filaments formed due to compressive shock waves tend to fragment sooner and have shorter lifetimes. For structures aligned radially along the axis of rotation in disk, the Type S formation scenario is likely to be the dominant mode, especially for quiescent environments. These filaments will be longer, more stable and have lifetimes that depend most on their own gas accretion rates. Filaments with alignments along spiral arms or perpendicular to the axis of rotation, on the other hand, are dominated by the compressive shock which will pile material into them and quickly promote fragmentation.

\begin{figure}
    \centering
    \includegraphics[width=1.0\linewidth]{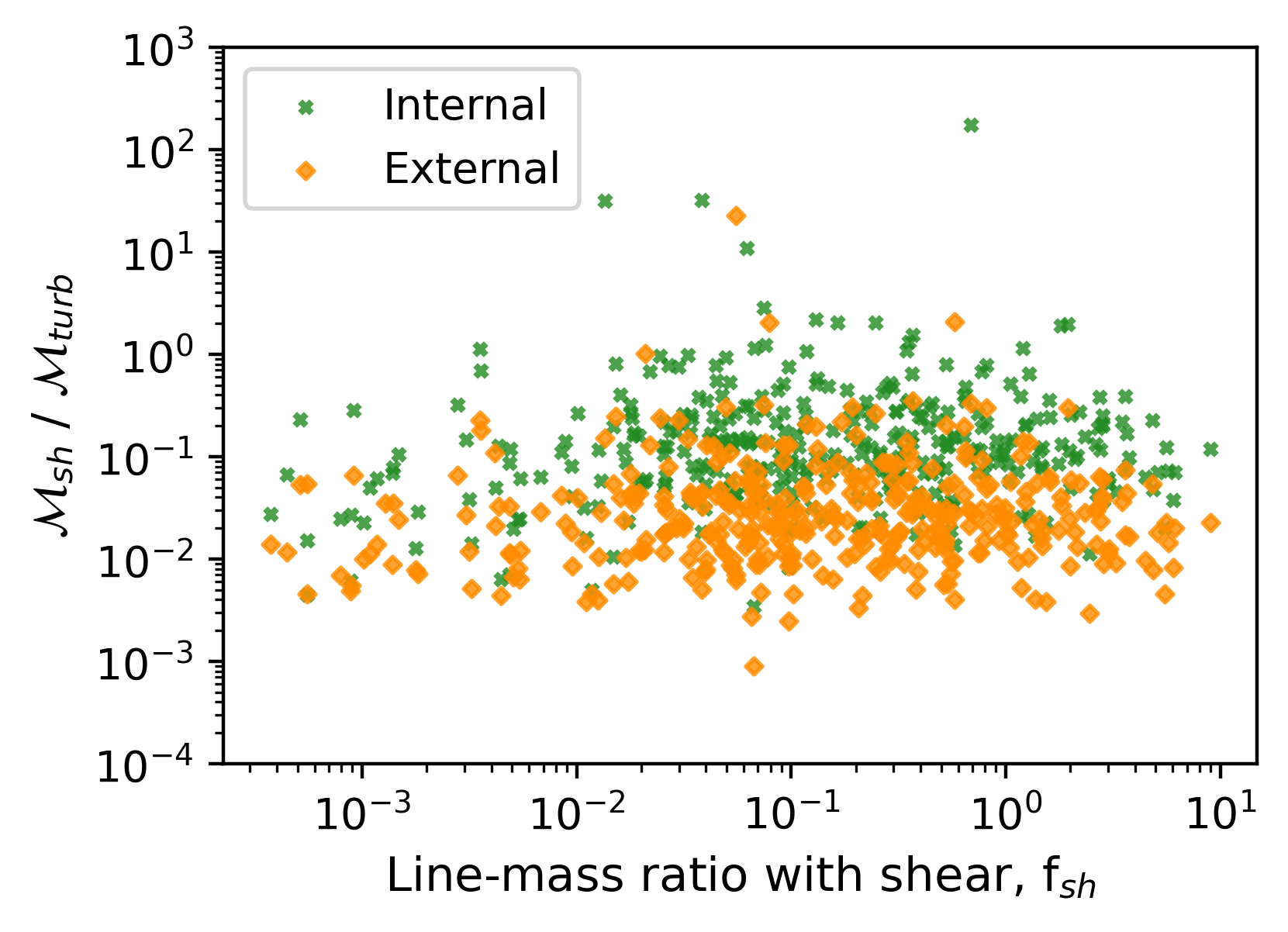}
    \caption{Comparison of shear and turbulent mach numbers vs. filament line mass ratio, measured from $f_{sh}$. Green X's show measurements from inside the filament, where perpendicular cuts have been limited to the width of the filament as measured by \texttt{FilFinder}. Orange diamonds are for measurements external to the filament width, up to a distance of 3 times the radius away on either side.}
    \label{fig:shear_turb_mach}
\end{figure}

In Figure \ref{fig:shear_turb_mach} we compare the influence of the shear internal and external to our filaments. We measure the ratio of a shear Mach number (similar to the turbulent Mach number, $\mathcal{M}_{sh}=\sigma_{sh}/c_s$) to the turbulent Mach number against the line-mass ratio with shear, $f_{sh}$. We measure shear Mach numbers both inside the bounds of the filament and external to it, up to a distance of triple the width of the filament. Since it operates on local scales throughout a galactic disk, shear can act both on the outside of the filament and inside of the structure. 

On average, internal Mach ratios are higher by an order of magnitude than outside the filaments. Given that there is no comparable rise in the turbulent velocity dispersions from outside to inside the filaments, this describes higher shear internally. The higher shear dispersion measured inside the filaments tell us that galactic shear, i.e. shear predominantly from spiral arms, does not play a large role in pulling the filament apart. The high internal shear in our filaments shows that shear is playing a more supportive role against gravitational collapse.

In summary, shear plays different roles based on where it is strongest. If shear is dominant external to the filament, it can act both as a destructive mechanism pulling the filament apart, as well as a supportive mechanism stretching a filament along its longitudinal axis preventing gravitational collapse. However, strong shear internal to a dense, self-gravitating filament would be aligned along its longitudinal axis. This might trigger a turbulent cascade acting during the formation of the filament and thereby supporting it against collapse (this is similarly discussed in \citet{HuWang2025}).\textbf{ }

We note that shear is expected to be anisotropic; however, in this work we are taking an average value. Recall from Figure \ref{fig:linemass_scatter} that the shear dispersion does not affect the number of supercritical filaments. Moreover, the ratios of shear to turbulent Mach numbers are constant across line mass ratios. Therefore, we conclude that the role of shear in most filaments is weaker compared to the effects of the magnetic field and turbulence inside the filament.

\subsection{Filamentary flows}\label{sec:filflows}
\begin{figure*}
    \centering
    \includegraphics[width=1.0\linewidth]{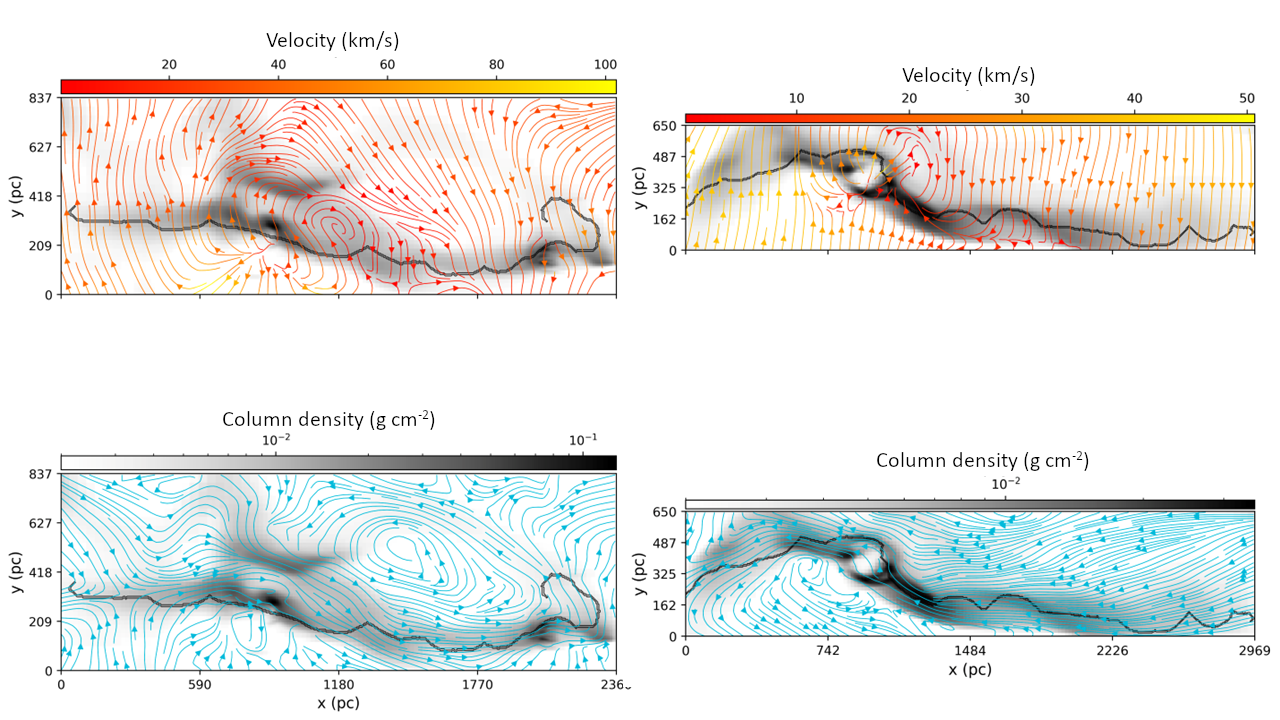}
    \caption{Column density maps for two filaments with large superbubbles expanding near the filament spine. In the top row we overplot the local gas velocity as streamlines in the red/yellow colormap. In the bottom row we show magnetic field morphology as the blue streamlines overplotted. The (x,y) axes for the filaments show the extent of the filament in each general direction and do not correspond to location in the galactic disk. The spines of the filaments are depicted via black contour.}
    \label{fig:filflows}
\end{figure*}

The velocity fields associated with filaments are of prime importance for understanding how filaments evolve and fragment. In Figure \ref{fig:filflows}, we show column density maps of 2 filaments chosen for the presence of large, active superbubbles in the filaments. Overlaid as streamlines we show the local velocity and the magnetic field morphology. In Appendix \ref{appa} we provide example filaments for other regions, including high rotation and quiet environments (Figures \ref{fig:youngbubs}-\ref{fig:rotate}). 

Interestingly, both filaments show gas flows that are largely perpendicular to the spine of the filament, whereas the magnetic fields tend to be parallel consistent with observations of similar gas column densities of $\sim7\times10^{21}-1\times10^{22}~\rm{cm}^{-2}$\citep{PlanckCollaborationAdam2016} (see Appendix \ref{appb} for the average orientations of our filaments). The most disruption to the magnetic field morphology happens where the bubbles bend the velocity field from their isotropic expansion. In both cases of Figure \ref{fig:filflows}, filaments show a change of flow direction near and within the bubbles as well. The flows diverge from a central point in the bubble and pile-up towards the high-density walls that form around them. These clumps coincide with the location of the bubble walls and show fragmentation generated by the compressive shock wave from an expanding supernova explosion.

Overall, gas flows around filaments tend to be misaligned with the magnetic field, especially for subcritical sections of the structure. Around supercritical pockets which display high column densities and evidence of inflowing gas, the magnetic and velocity fields align parallel to one another. We surmise that this happens primarily through the gravitational influence of the high column density gas on the magnetic field morphology. 

Moreover, within the bounds of the filaments, gas flows are ordered, and show low turbulence within the filaments consistent with dispersions around 10 km/s. However, we note that at a working resolution of $5.2~\rm{pc}$, the internal structures of our filaments are not well resolved. Further work at higher resolutions, thereby resolving the internal structure of filaments, will be important to better compute the magnitude of the internal filament shearing motions. 

Finally, for filaments in sufficiently quiet backgrounds (in our case, those shown in Figures \ref{fig:filflows},\ref{fig:youngbubs}, \ref{fig:nothing} \& \ref{fig:rotate}) there is a net rotation of the structures on a large-scale that is distinct from the overall rotation of the galactic disk. This is not the case for filaments shown in Figure \ref{fig:squeeze}, which have multiple bubbles in their vicinity and therefore significant turbulent mixing. This morphology in the velocity fields shows that most filaments may form from obliquely colliding shock waves which impart a net rotation from their glancing collisions.

\subsection{Dependence on Galactocentric Radius}
\begin{figure}
    \centering
    \includegraphics[width=1.0\linewidth]{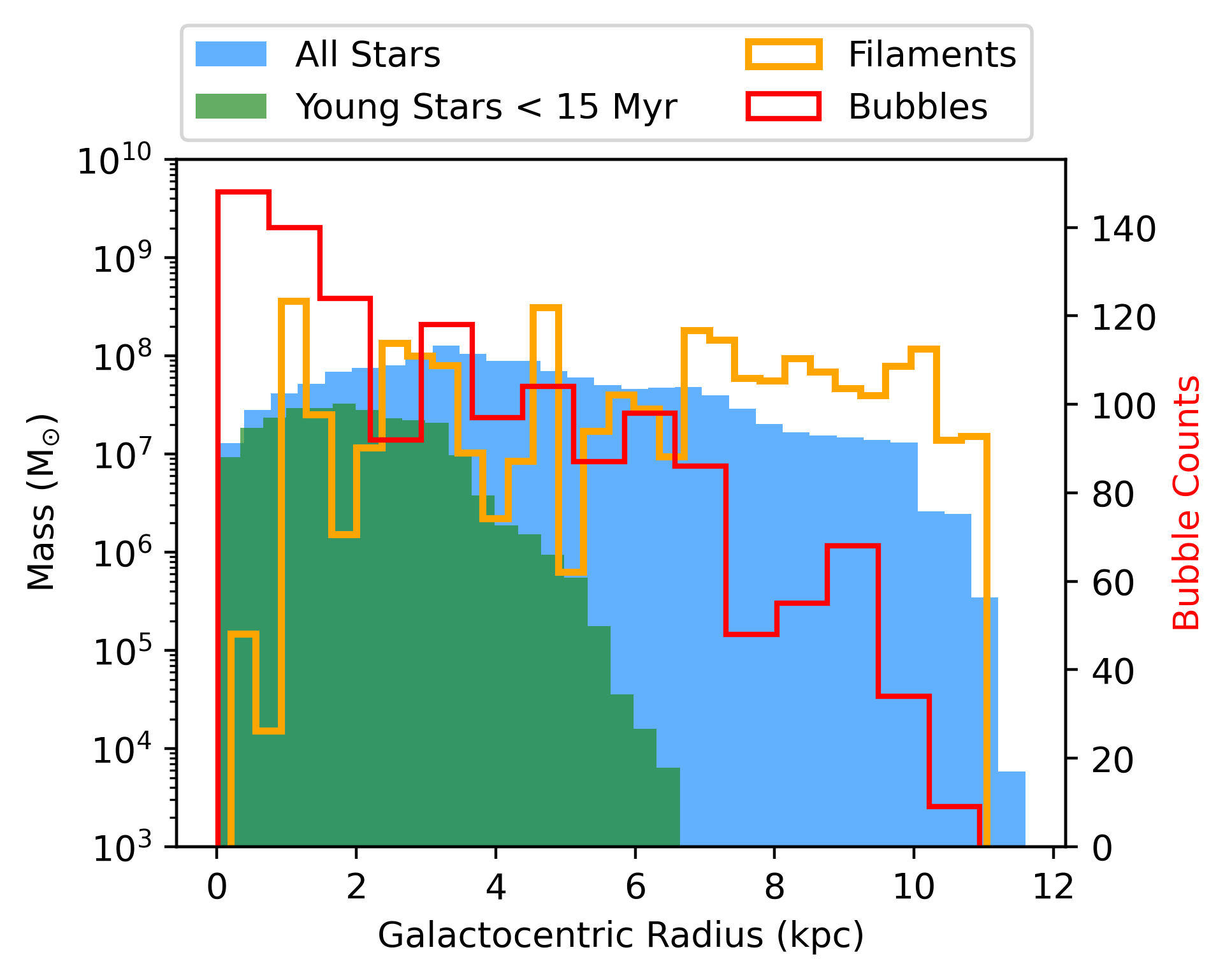}
    \caption{Mass distributions vs. galactocentric radius for young stars in our disk less than 15 Myr in green, in blue all the stars in our disk and in the orange line the filaments making up our galactic disk. On the righthand y-axis, we give the bubble counts per radial bin, which is depicted as the red line in the plot.}
    \label{fig:starmasshisto}
\end{figure}

In Figure \ref{fig:starmasshisto}, we plot the mass distributions of the filaments and stars in our simulation as a function of galactocentric radius and the counts of superbubbles across galactocentric radii. We count the position of a bubble in the galactic disk as the average GC radius across all the cells part of that bubble. From this, we see that the filamentary mass is approximately uniform across galactic radii. On the other hand, the young stars and the majority of the bubbles exist at inner radii less than 6 kpc. 

These differences provide an interesting picture of how active star formation and filament fragmentation in the galaxy are typically distributed, given that our simulation has reached a steady global star formation rate. Most of the very young stars have formed in the inner radii of the galaxy, which suggests that the supercritical  filaments there undergo rather rapid fragmentation into stars. As such, the bulk of the filamentary structure in this inner region of the disk will be either newly forming  as a consequence of the higher shock rates from spiral arms and superbubbles, or that will dissipate before becoming supercritical. We find it most likely that of these two options, the majority of the filamentary structure must be in the state of actively forming due to the increased counts of superbubbles in this region of the disk, which contribute to high levels of turbulence and active mixing of the gas here.

On the other hand, the filaments in outer regions of the galactic disk are essentially devoid of young stellar populations; only stars older than 15 Myr are found. These outer radii have higher gravitational support and a flatter rotation curve, such that shear in the outer disk could prevent these structures from collapsing for up to hundreds of Myr \citep{JeffresonKruijssen2018}. However, the much shorter 10-20 Myr for the onset of supernovae will dominate the lifetimes of the filaments. In fact, the low bubble counts at outer radii combined with the presence of stars older than 15 Myr supports that much of the filamentary structure must be new, and that the destruction of structures by supernovae has not yet begun. These structures therefore must be approaching a state of collapse. As such the outer radii of the galactic disk, which host the longer filamentary structures, will at this point also preferentially host the supercritical filaments which have begun their fragmentation into star-forming regions. 

\subsection{Superbubble proximity and external pressure}\label{sec:pressure}
In general, external pressure can squeeze filaments, and for non-isothermal filaments at least, this can push them towards criticality \citep{Andre2014}. Using our bubble map, we investigate the role that superbubbles and external pressures play on the line masses and fragmentation of filaments. To do so, we calculate the nearness of a bubble to a filament by calculating the distance to all bubbles in the vicinity of each filament in our data. The distance is defined as the distance from the filament to the nearest edge of the bubble. 

\begin{figure}
    \centering
    \includegraphics[width=1.0\linewidth]{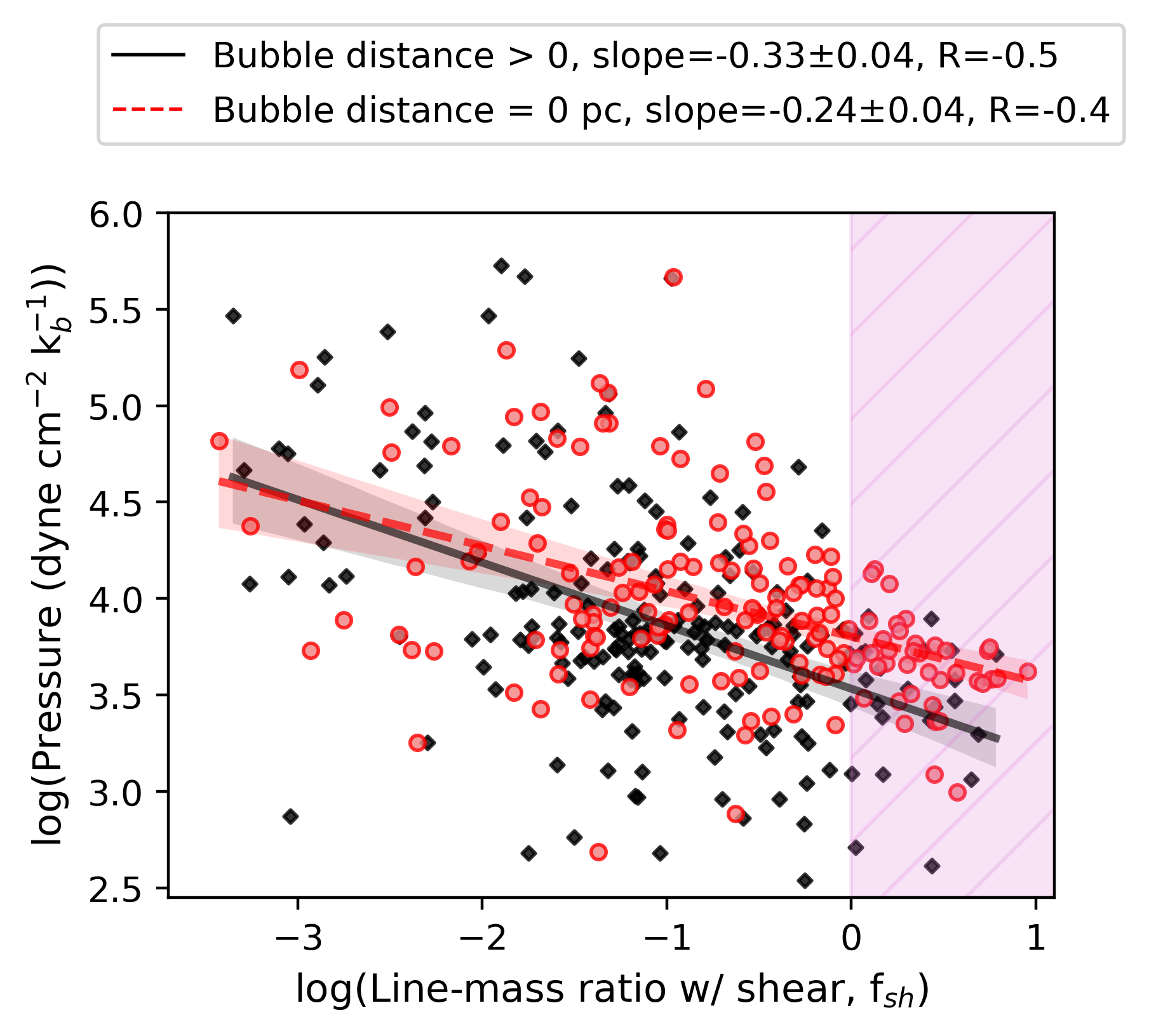}
    \caption{Surface pressure versus the shear line mass ratio of the filaments. The purple shaded region denotes the supercritical region -- line mass ratios $>1$. Filaments where superbubbles are coincident with the structure (i.e. have a distance of 0 pc) are shown as red circles. The black diamonds show the filaments with bubbles greater than 0 pc away. For both sets of data we plot the linear regression fits to the points.}
    \label{fig:pres_lm_bub}
\end{figure}

In Figure \ref{fig:pres_lm_bub}, we show the relation between the surface pressure of the filaments and their line mass ratio. The surface pressure is calculated as the total pressure, with thermal, turbulent and magnetic contributions, as measured along the edges of the filament as defined by its width measured via \texttt{FilFinder}. We fit the relations to a line, with the filaments that touch a bubble shown in red. We see no significant change in the relationship between surface pressure and line mass ratio between the population of filaments not touching bubbles and just those that overlap with bubbles. The high scatter in surface pressures, especially for the subcritical filaments, makes any relationship very weak. 

In order to probe any connection between surface pressure on filaments, their internal structure, and stability, we introduce the pressure ratio: 

\begin{equation}
    \chi_f = P_{surf}/P_{central}, 
\end{equation}    
\noindent where $P_{surf}$ represents the pressure measured at the surface of the filament and $P_{central}$ the pressure measured along the dense, central ridge. The motivation here is that if this ratio is less than unity, then the self-gravity of the filament must dominate the filament structure. It accounts for the observed radial profiles of self gravitating filaments as in \citet{Ostriker1964,FiegePudritz2000}. This is also motivated by considerations of the vertical pressure exerted by the weight of the gas on the galactic midplane \citep{KimOstriker2015}. 

This ratio $\chi_f$ usefully describes the different pressure regimes in the galaxy and their effects on filament structure. For values $\chi_f > 1.0$ the filament experiences higher surface pressures than central pressures and is compressed by the gas around it. For example, a filament with $\chi_f > 1.0$ may be a young filament being formed in the swept up  shell of an expanding supernova bubble. 

On the other hand, values of $\chi_f \leq 1.0$ describe filaments whose central pressures are much higher than their surface pressures. This describes a scenario in which the weight of all the gas on the filament is pushing on the central ridge and compressing it. A filament in this regime may be either subcritical or supercritical, depending on whether it has reached the point where its internal motions can no longer support it or not. In the former scenario, a subcritical but low $\chi_f$ filament will be fully formed and self-gravitating. It has a dense ridge, but has yet to begin fragmenting or collapsing along this ridge. In the latter, a supercritical filament must always have a low $\chi_f$; a fragmenting filament sees higher central pressure than on its surface, as the interior has begun collapsing into the spine allowing the surface to fall in.

\begin{figure}
    \centering
    \includegraphics[width=1.0\linewidth]{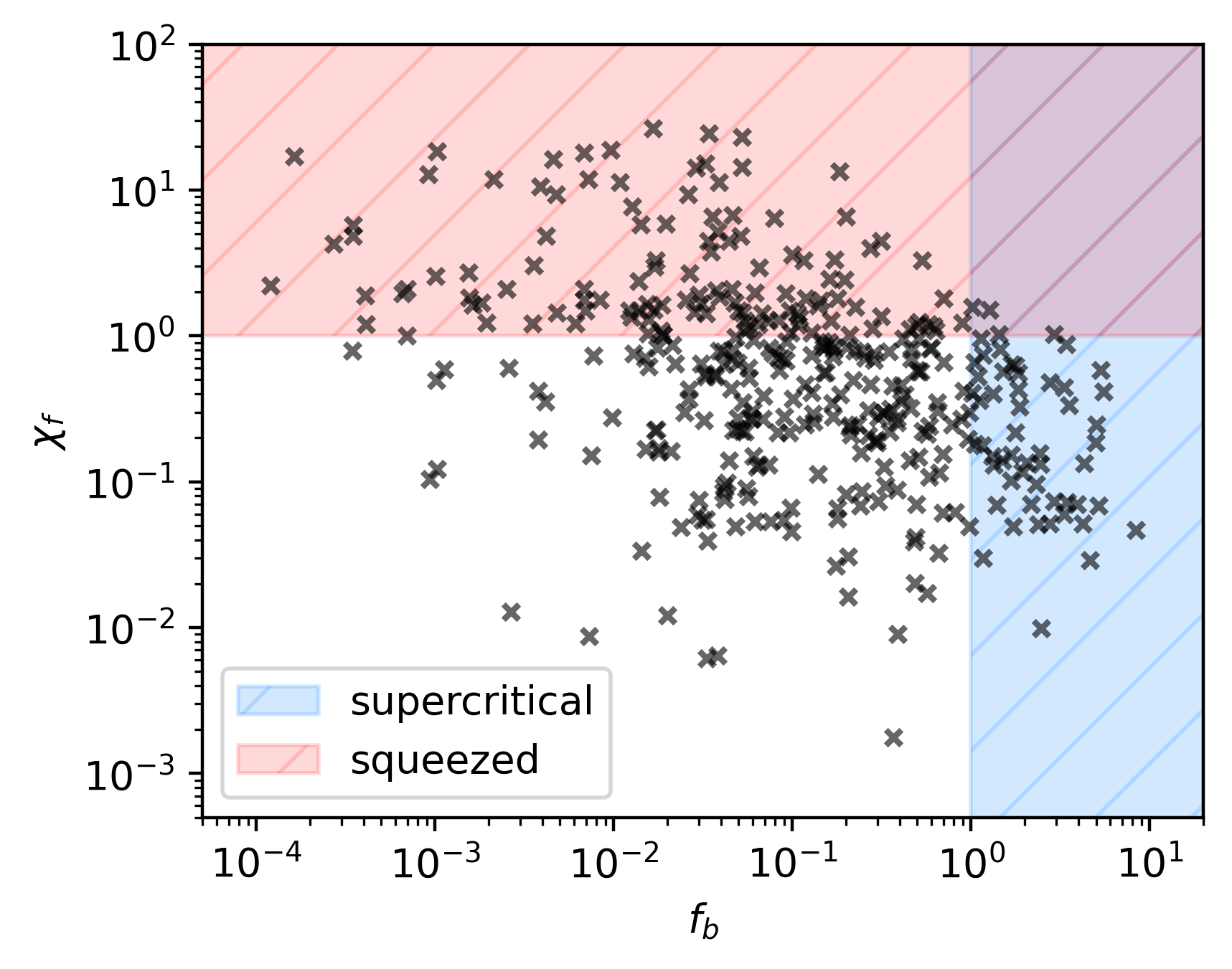}
    \caption{Filament pressure ratio, $\chi_f$, versus the line mass ratio of the filaments. The blue shaded region denotes line mass ratios greater than 1 (supercritical filaments). The red shaded region denotes pressure ratios greater than 1.}
    \label{fig:chi}
\end{figure}

In Figure \ref{fig:chi}, we show the relationship between $\chi_f$ and the line-mass ratio for each of our filaments. The shaded blue is the supercritical regime, and shaded red is the low $\chi_f$ regime. One filament exists in the combination of these regimes but is close to the critical line-mass ratio that could result from a small variation in the measured mass or length. Save this one filament, all supercritical filaments exist in a supercritical, low $\chi_f$ regime, describing their collapse and fragmentation. On the other hand, the subcritical population is indeed separate into two areas. The filaments we suspect to be actively forming exist in the red shaded region, whereas the truly subcritical filaments exist in this transient regime of the plot. 

\begin{figure*}
    \centering
    \includegraphics[width=1.0\linewidth]{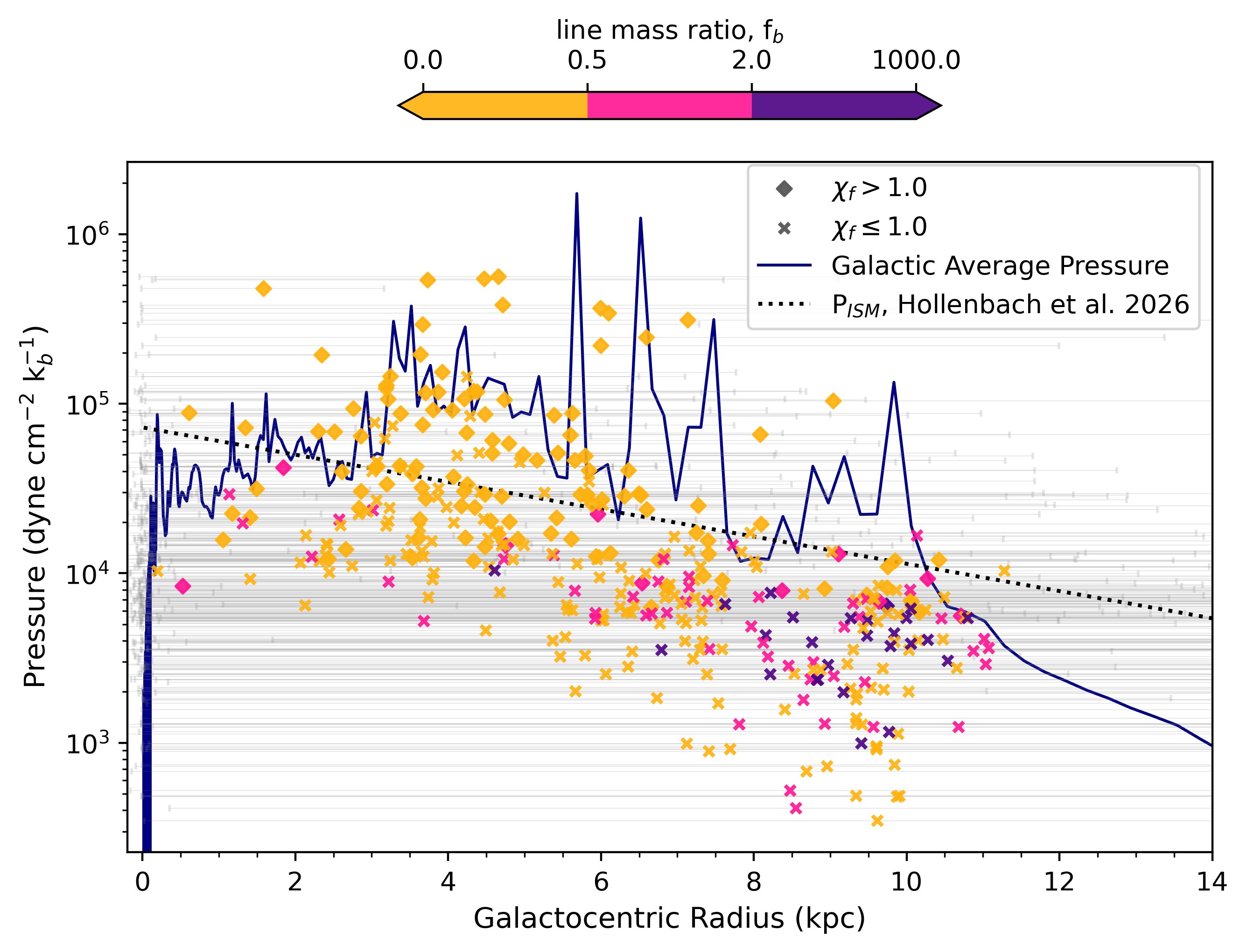}
    \caption{Total surface pressure of filaments vs. galactocentric radius. Total pressure includes thermal, turbulent and magnetic pressures of the gas. Error bars on the galactocentric radius show the minimum and maximum radius values, with the marker being placed at the average radius of the filament position. We represent $\chi_f > 1.0$ with diamonds, and $\chi_f \leq 1.0$ with x's. The colour bar depicts the line mass ratio of the filament, with gold being subcritical, pink being transcritical and purple being supercritical. We plot the azimuthal averaged galactic pressure in the navy line. The dotted black line shows the ISM scaling relation from \citet{HollenbachParravano2026}.}
    \label{fig:net_gcrad}
\end{figure*}

In Figure \ref{fig:net_gcrad}, we show the surface pressure of our filaments versus their average galactocentric radius position. We separate the filaments in three categories by colour: subcritical filaments, representing ratios less than 0.5; transcritical filaments with ratios between 0.5 and 2.0; and supercritical filaments with ratios above 2.0. In the same plot, we azimuthally average the galactic pressure for the height of the entire disk -- instead of just the midplane -- across radial bins. 

From this, we can see that all but one of our supercritical filaments as well as the majority of the transcritical filaments are in the regime $\chi_f \leq 1.0$. There is a significant population of subcritical filaments marked with x's, which are the true subcritical filaments identified with $\chi_f\leq 1.0$. On the other hand, the majority of the diamond markers also have subcritical line mass ratios supporting the idea that $\chi_f>1.0$ describes a transient regime in which filaments are not yet fully formed, self-gravitating structures and therefore cannot be supercritical. Additionally, the high $\chi_f$ filaments are clustered towards the inner galactic disk. Moreover, it is notable that the average surface pressures of the filaments, approximately $10^4$ dyne cm$^{-2}$ k$_b^{-1}$ at Solar neighbourhood, tend towards lower than galactic average pressures by an order of magnitude. This is especially true for supercritical filaments with $\chi_f\le1.0$, as their higher central pressures also imply lower surface pressures. These filaments also lie below the midplane scaling relation from \citet{HollenbachParravano2026}. The subcritical filaments with $\chi_f>1.0$ follow both the midplane scaling from \citet{HollenbachParravano2026} and our galactic average pressure well. We can conclude that subcritical filaments with higher surface pressures (i.e. those filaments which are likely still forming) are more tightly correlated with the pressure in the galactic disk than the supercritical filaments. There is also a strong link with galactocentric radius for these filament populations, suggesting that supercritical filaments at outer radii are less likely to be in the midplane of the disk and may sit at higher vertical positions. 

As noted for Figure \ref{fig:starmasshisto}, the inner radii host all the young stars and the majority of the superbubbles in our galactic disk. This highlights that many supernovae have recently gone off at these inner radii and produced a highly turbulent ISM. The mixing and shockwave propagation from this turbulent acts to break up some existing filaments and plow the gas into new filaments. This is consistent with being in a high $\chi_f$ regime and with overall higher surface pressures.

The supercritical filaments and the majority of the low $\chi_f$ regime filaments exist at outer radii of the galactic disk. These filaments have either not quite accreted enough gas to become supercritical or are already supercritical and actively collapsing. However, as we have shown in \citetalias{PillsworthRoscoe2025a}, these average line-mass ratios, and now the average $\chi_f$ values, may not always accurately describe the fragmentation state of a filament. For filaments much longer than 100 pc, fluctuations in the line mass can drive local fragmentation. It is likely that some very long subcritical filaments have localized pockets of fragmentation which will not be significant enough to affect the average. We expect the scenarios to still only occur for the filaments in our new ``true'' subcritical regime, but future work will be needed to investigate the local role of $\chi_f$. 

\section{Summary and Conclusions}\label{conclusions}
In this paper we have analyzed both the filamentary and superbubble structures of a Milky Way-like galaxy simulated in \textsc{ramses} and their behaviour in the dynamical environment of the galaxy. \textbf{ }In order to better compare the results of our simulations with observations, we first developed and applied several data processing techniques to our simulations, including the bubble identification code \texttt{perch}, the application of column density thresholds, and projection effects. Comparing the results with our previous work \citetalias{PillsworthRoscoe2025a}, we find that it is important to limit the projection depth to exclude the hot gas ejected into the galactic halo resulting from galactic fountain flows. When using face-on full disk projections, the hot gas makes an apparent contribution to filament properties that are not real. These limited depths are also important to sharpen width measurements for kpc length filaments, as the exclusion of hot, diffuse gas helps clarify the locations of these very long structures. 

\textbf{ }We then applied these sharper analytical tools, such as our preferred ``pillbox'' dataset (Table \ref{table1}), to investigate the effects of dynamical processes in galaxies on filament properties and fragmentation. We found an abundance of new results including the relatively minor role of shear in filament formation and fragmentation, the importance of magnetic field contributions to the critical line mass and low frequency of supercritical filaments, and the variation of filament fragmentation and formation as a function of galactic radius. The latter effect seems to be closely related to the variation in galactic ISM pressure.

We summarize our conclusions as follows: 

\begin{enumerate}
    \item We find that filaments comprise the majority of the galaxy's gas mass budget (78\% of the total gas mass) in accordance with \citet{HacarClark2023}. We further find that filamentary structures make up a small fraction of the overall surface area of a galactic disk -- 49 kpc$^2$ -- versus the 200 kpc$^2$ that superbubbles occupy. These are analogous to 3D volume filling factors and show that the majority of the star-forming gas indeed occupies very little of the physical space in a galaxy.

    \item Magnetic fields are undeniably important to the critical support a filament experiences. The inclusion of magnetic corrections to the critical line mass calculation shifts the percentage of supercritical filaments by number to 15\%, much closer inline with star formation efficiencies in our simulated galaxy of 10\%. We recommend that further filament studies should include magnetic field corrections to the critical line masses at minimum in order to fully understand the fragmentation and star formation within filamentary structures.

    \item Magnetic field morphologies of filaments tend to be parallel on large kpc scales, and misaligned with velocity flows. 
    
    \item Shear dispersions external to the filament have little effect on the support of the filamentary structure. However, on average filaments in our simulated galaxy experience higher internal shear dispersions than external. Thus, while shear may have little effect on the environment-level support for a filament, the internal motions of a filament in preventing collapse and fragmentation may in part come from shearing motions.

    \item Superbubble locations in the galactic disk seem to trace the young stars of the disk. This matches with recent \textit{JWST} observations of nearby galaxies that suggest bubbles trace star formation in a galactic disk \citep[e.g.,][]{WatkinsBarnes2023}.

    \item Low-density voids and superbubbles do not seem to dictate the locations or properties of filaments. There is no correlation between bubble proximity and filament supercriticality or surface pressure.

    \item The ratio of the surface to central pressure of a filament, $\chi_f$, is an important parameter in describing its evolutionary state. We find that whether $\chi_f$ is greater or smaller than unity can help differentiate between still forming filaments in which the gas is still being squeezed into a filament, and  trans or supercritical filaments which  will  begin or are fragmenting.

    \item The pressures that filament undergo in the ISM vary across the galaxy. Higher surface pressures (and therefore higher $\chi_f$ measures) correlate with filaments at smaller galactocentric radii, which trace the youngest stars of the disk. On the other hand, supercritical filaments which have yet to begin forming stars exist in outer radii where only old stars exist and have lower pressures bound by the mid-plane disk pressure.
\end{enumerate}

\begin{acknowledgements}
The authors thank an anonymous referee whose report was very useful in clarifying the manuscript. We would also like to thank Ralf Klessen, James Wadsley, Claude Cournoyer-Cloutier and Jeremy Karam for insightful discussions during the preparation of this work. RP acknowledges support from an NSERC CGS-D research scholarship. REP acknowledges funding support from an NSERC Discovery Grant. The computational resources for this project were enabled by a grant to REP from Compute Canada/Digital Alliance Canada and carried out on the Cedar/Fir computing cluster.
\end{acknowledgements}

\software{matplotlib \citep{Hunter2007}, YT \cite{TurkSmith2011}, astropy \citep{AstropyCollaborationRobitaille2013}, scipy \citep{VirtanenGommers2020}, powerlaw \citep{AlstottBullmore2014}, filfinder \citep{KochRosolowsky2015}, perch (O'Neill et al., in preparation)}

\bibliographystyle{aasjournal}
\bibliography{PhDpaper2_fildynamics}

\appendix

\section{Filament Internals}\label{appa}
\begin{figure}
    \centering
    \includegraphics[width=0.8\linewidth]{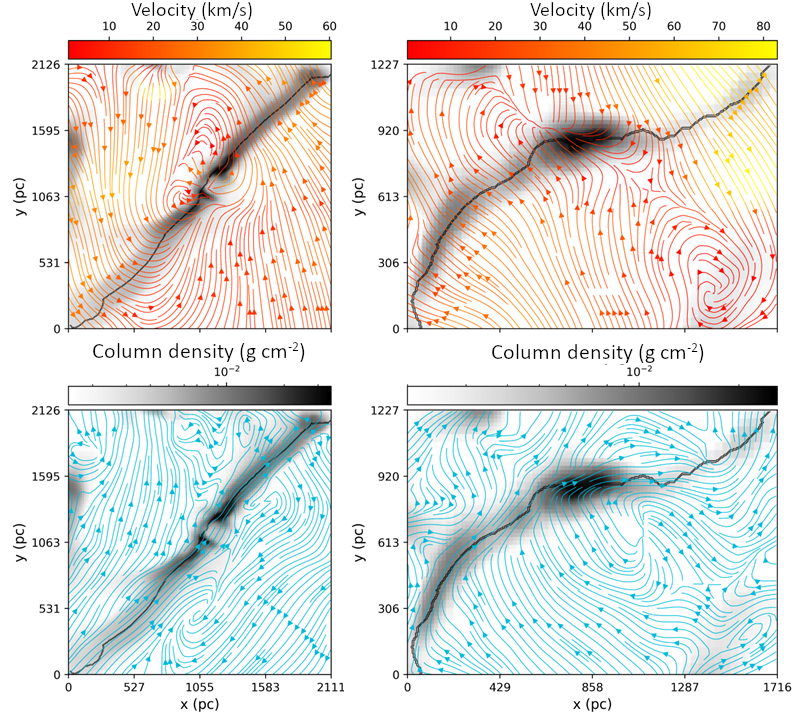}
    \caption{Same as Figure \ref{fig:filflows}, but for two filaments with smaller, young superbubbles expanding near the filament spine.}
    \label{fig:youngbubs}
\end{figure}

\begin{figure}
    \centering
    \includegraphics[width=0.85\linewidth]{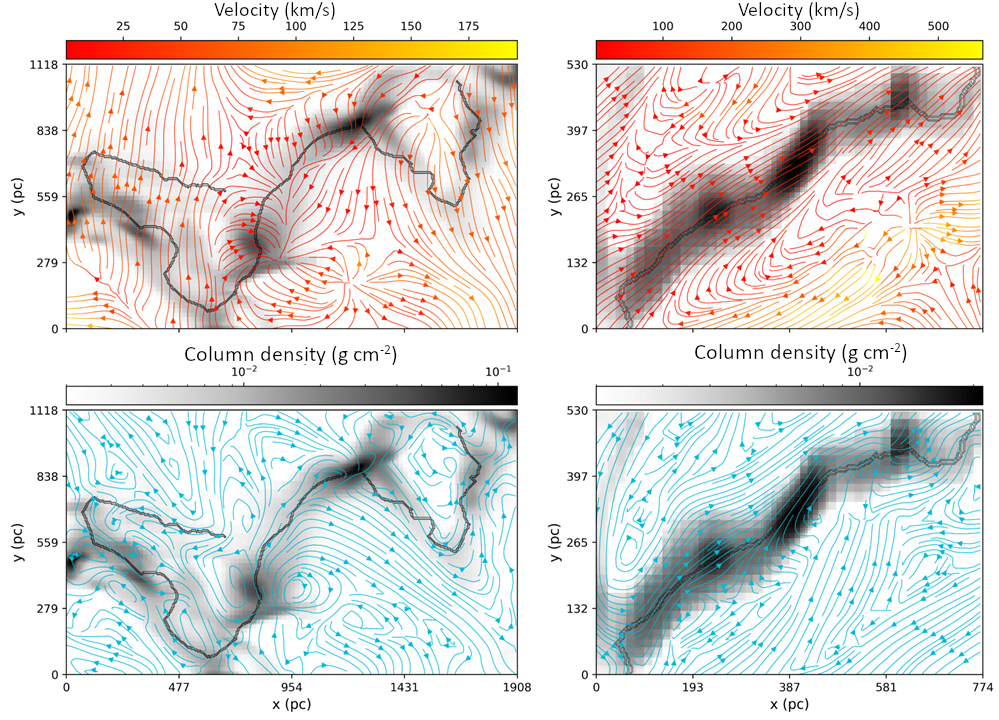}
    \caption{Same as Figure \ref{fig:filflows}, but for two filaments with many superbubbles expanding around the filaments on both sides of the structure.}
    \label{fig:squeeze}
\end{figure}

\begin{figure}
    \centering
    \includegraphics[width=0.85\linewidth]{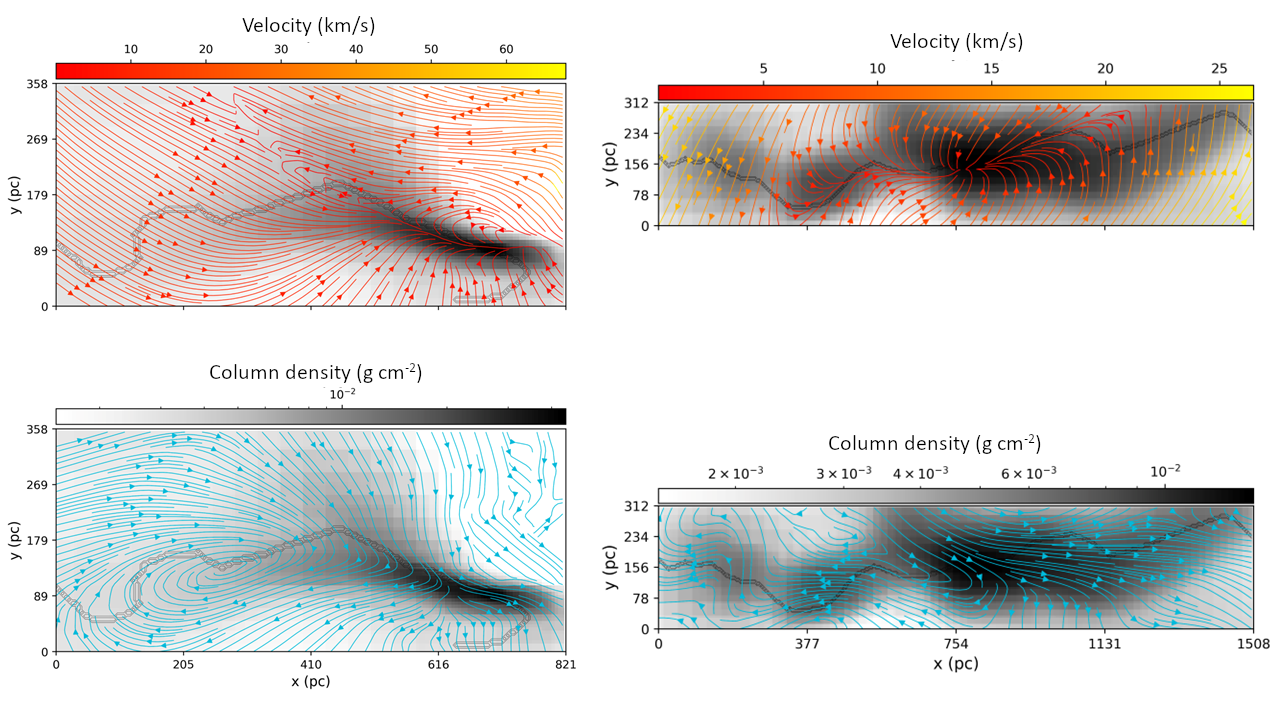}
    \caption{Same as Figure \ref{fig:filflows}, but for two filaments with relatively quiet environments and no superbubbles nearby.}
    \label{fig:nothing}
\end{figure}

\begin{figure}
    \centering
    \includegraphics[width=0.8\linewidth]{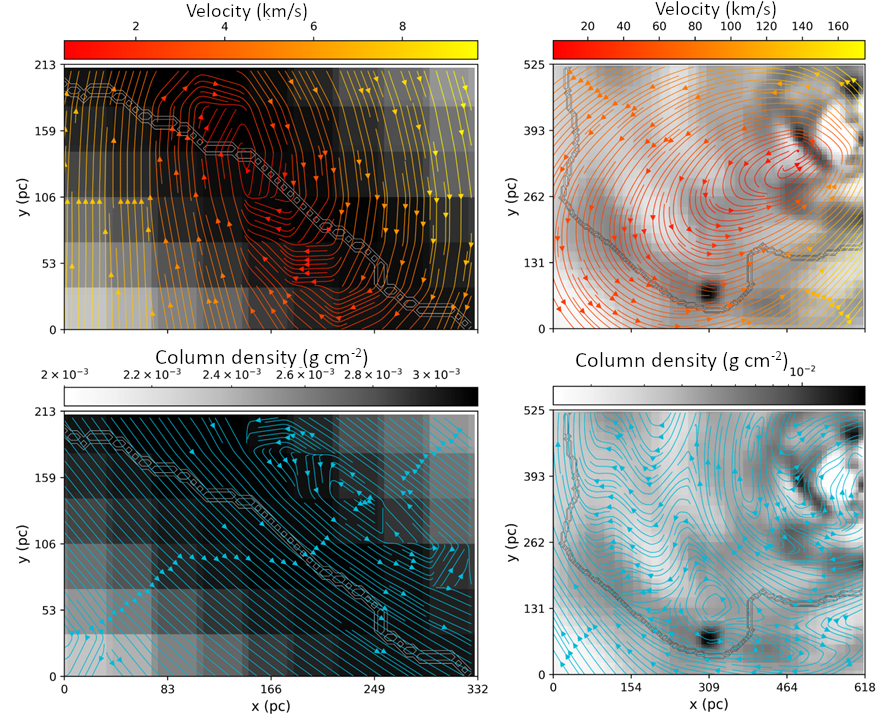}
    \caption{Same as Figure \ref{fig:filflows}, but for two filaments which show extreme net rotations.}
    \label{fig:rotate}
\end{figure}

Here we provide the figures for other filaments discussed in \S \ref{sec:filflows}. In Figures \ref{fig:youngbubs}-\ref{fig:rotate}, we show the velocity flows and magnetic field morphologies overplotted on the column density maps of filaments with different local environments. We show two filaments each for environments with young, hot, low-density voids (Figure \ref{fig:youngbubs}), environments with many voids which squeeze the filament (Figure \ref{fig:squeeze}), quiescent environments with no voids nearby (Figure \ref{fig:nothing}), and environments with large net rotation (Figure \ref{fig:rotate}).

\section{Magnetic Field Orientations}\label{appb}
\begin{figure}
    \centering
    \includegraphics[width=0.7\linewidth]{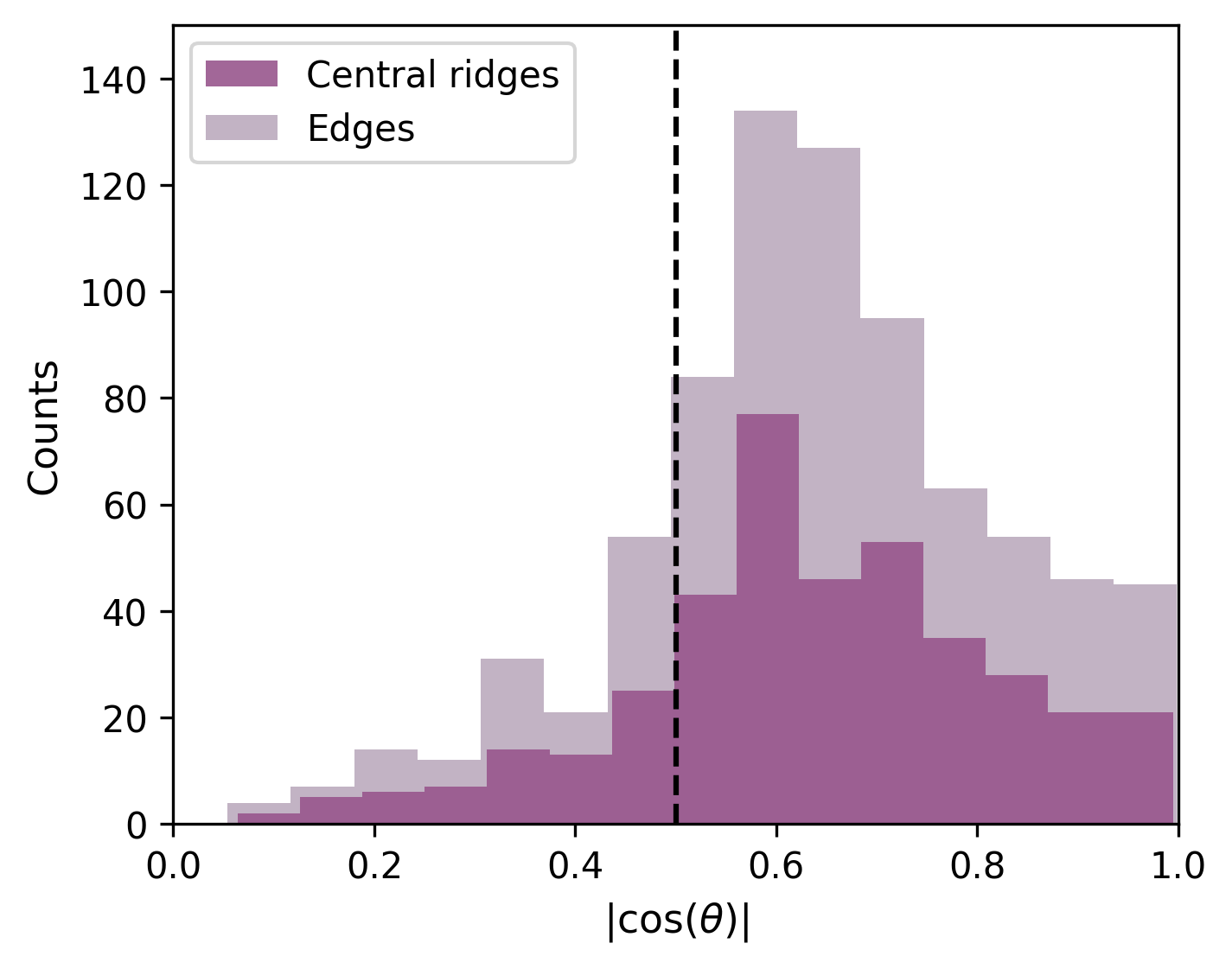}
    \caption{Histogram of relative orientations between the filaments and the magnetic field vectors. In the lighter purple we show measurements for the dense, central ridges (or spines) of the filaments. In the darker purple, we show the measurements along the edges of the structures. The vertical dashed line delineates between parallel measurements to the left ($|cos(\theta)|<0.5$) and perpendicular measurements to the right ($|cos(\theta)|>0.5$).}
    \label{fig:all_bori}
\end{figure}

Magnetic fields overall have largely smooth morphologies in and around filamentary structures. In Figure \ref{fig:all_bori}, we show the distributions of average relative orientations between magnetic field vectors and the filament both for the central dense ridge and the edges of the structures. We find that the majority of filaments have largely parallel orientations, consistent with results in the CNM \citepalias{PillsworthPudritz2024b}, which is the density regime our filaments reside in. Few filaments have perpendicular orientations (defined as $cos(\theta)>0.5$) on average, but the central ridges have slightly higher perpendicular orientation occurrences than the edge measurements. The smooth (often parallel) magnetic field acts to support filaments, mirroring the large shift in supercritical percentages discussed in \S \ref{sec:linemass}. 

\end{document}